\documentclass[twocolumn,twocolappendix]{aastex63}

\newcommand{\fcaglp}{\affiliation{Facultad de Ciencias Astron\'omicas y Geof\'{\i}sicas, Universidad Nacional de La Plata, Paseo del Bosque, B1900FWA La Plata, Argentina}}
\newcommand{\iar}{\affiliation{Instituto Argentino de Radioastronom\'ia (CCT La Plata, CONICET; CICPBA; UNLP),\\ C.C.5, (1894) Villa Elisa, Buenos Aires, Argentina.}}
\newcommand{\roch}{\affiliation{Center for Computational Relativity and Gravitation, School of Mathematical Sciences, Rochester Institute of Technology, 85 Lomb Memorial Drive, Rochester,New York 14623, USA}}
\newcommand{\holanda}{\affiliation{Kapteyn Astronomical Institute, University of Groningen, P.O. BOX 800, 9700 AV Groningen, The Netherlands}}
\usepackage{graphicx}
\usepackage{gensymb}
\usepackage{amsmath}
\usepackage{float}

\received{October 5, 2020}
\revised{November 24, 2020}
\accepted{November 25, 2020}
\submitjournal{ApJ}

\shorttitle{IAR timing analysis of PSR$~$J0437$-$4715}
\shortauthors{Sosa Fiscella et al.}
\graphicspath{{./}{figures/}}

\begin{document}

\title{PSR~J0437$-$4715: The Argentine Institute of Radioastronomy 2019--2020 Observational Campaign}

\title{PSR~J0437$-$4715: The Argentine Institute of Radioastronomy \\~~~~~~~~~~~~~~~~~~~~2019-2020 Observational Campaign}

    \author[0000-0002-5176-2924]{V.~Sosa~Fiscella} \fcaglp \iar
    \author[0000-0002-5761-2417]{S.~del~Palacio} \iar
    \author[0000-0002-5427-1207]{L.~Combi} \iar \roch
    \author[0000-0002-6400-9640]{C.~O.~Lousto} \roch
    \author{J.~A.~Combi} \fcaglp \iar
    \author{G. Gancio} \iar
    \author[0000-0001-9072-4069]{F.~Garc\'ia} \iar \holanda
    \author[0000-0001-7941-801X]{E.~Guti\'errez} \iar
    \author{F.~Hauscarriaga} \iar
    \author{P.~Kornecki} \iar
    \author{F.~G.~L\'opez Armengol} \roch
    \author{G.~C.~Mancuso} \iar
    \author{A.~L.~M\"uller} \iar
    \author{A.~Simaz~Bunzel} \iar

\correspondingauthor{C.~O.~Lousto} \email{colsma@rit.edu}



\begin{abstract}
We present the first-year data set of high-cadence, long-duration observations of the bright millisecond pulsar J0437$-$4715 obtained in the Argentine Institute of Radioastronomy (IAR). Using two single-dish 30 m radio antennas, we gather more than 700 hr of good-quality data with timing precision better than 1~$\mu$s. We characterize the white and red timing noise in IAR's observations, we quantify the effects of scintillation, and we perform single-pulsar searches of continuous gravitational waves, setting constraints in the nHz--$\mu$Hz frequency range. We demonstrate IAR's potential for performing pulsar monitoring in the 1.4 GHz radio band for long periods of time with a daily cadence. In particular, we conclude that the ongoing observational campaign of J0437$-$4715 can contribute to increase the sensitivity of the existing pulsar-timing arrays.
\end{abstract}



%
\section{Introduction\label{sec:introduction}}
%

The Argentine Institute of Radioastronomy (IAR) is equipped with two single-dish 30 meter antennas--dubbed A1 and A2\footnote{In 2019 the antennas A1 and A2 were renamed ``Varsavsky'' and ``Bajaja'', respectively.}--capable of performing daily observations of pulsars in the southern hemisphere at 1.4 GHz. These antennas were recently refurbished to obtain high-quality timing observations as described in \cite{Gancio2020}. 

Pulsar Monitoring in Argentina\footnote{\url{http://puma.iar.unlp.edu.ar}} (PuMA), is a scientific collaboration dedicated to pulsar observations from the southern hemisphere.
As part of IAR's observatory developing stage, accurate timing observations of the millisecond pulsar (MSP) J0437$-$4715 with both antennas 
have been carried out since 2019 April 22, with a daily follow-up only interrupted during hardware upgrades or bad weather conditions. 

The MSP J0437$-$4715 was discovered in 1993 by \cite{Johnston1993}, and it is one of the brightest (mean flux density $S_{1400}=150.2$~mJy) and closest 
($d=156.79 \pm 0.25$~pc) pulsars. It has a short period ($P=5.758$~ms) and it is one of the most massive pulsars known to date \citep[$m = 1.44 \pm 0.07~\mathrm{M_\odot}$;][]{Reardon2018}. This pulsar is in a binary system and in an almost circular orbit of period 5.74 days. The secondary star is a low-mass ($\sim0.2~\mathrm{M_\odot}$) helium white dwarf with strong visible emission \citep{Danziger1993}. In the interstellar region, an optical bow shock was also reported by \cite{Bell1993}. In addition, it was the first MSP detected in X-rays \citep{Becker1993} and the only one for which individual pulses have been studied. It is also the first one detected in the ultraviolet, although in this wavelength its spectrum is consistent with that of a blackbody \citep{lorimer2012handbook} and pulsed emission was not observed \citep{Kargaltsev2004}. 

Because of its proximity to Earth, J0437$-$4715 is one of the two pulsars with a well-determined three-dimensional orientation of the orbit \citep{vanStraten2001}. In addition, its radio emission does not present much nulling, short-scale variation of its integrated profile, or mode-changing \citep{Vivekanand1998}, phenomena associated with longer-period pulsars. This suggests that the origin of the radiative processes in this pulsar is different from the mechanisms in regular pulsars. Moreover, J0437$-$4715 displays intrinsic and quasiperiodic variations in its flux, \citep[not observed in other pulsars;][]{Vivekanand1998}, and extrinsic variations, due to interstellar medium (ISM) scintillations \citep{Oslowski2014}.

J0437$-$4715 stands out for having an extremely stable rotation rate which makes it a natural clock with a similar stability to that of an atomic clock \citep{Hartnett2011} and better over timescales longer than a year \citep{Matsakis1997}. Only two other pulsars, PSR B1855+09 and PSR B1937+21, have a comparable stability \citep{1994ApJ...428..713K}. These characteristics of J0437$-$4715 make it an ideal candidate for pulsar-timing studies. Its high declination in the southern hemisphere makes its observation from the northern hemisphere difficult to achieve \citep[see Fig.~2 of][]{Ferdman2010}. This MSP is also in the opposite direction to the Galactic center, where few pulsars are observed. For these reasons, performing daily observations of J0437$-$4715 is a key science project at IAR, improving upon the weekly to monthly cadence of other observatories in the International Pulsar Timing Array consortium \citep[IPTA;][]{Perera2019, Lam2020}. This is currently of particular interest as the NANOGrav collaboration is on the verge of detecting an isotropic stochastic gravitational wave background~\citep[GWB;][]{Arzoumanian:2020vkk}.

In this work, we use the properties of J0437$-$4715--high rotational stability, high luminosity, and short period--to assess the quality of the observations at IAR with both antennas. This builds upon the preliminary analysis presented in \cite{Gancio2020} which suggested they reach a precision of $\lesssim 1~\mu$s.

The paper is organized as follows: Sect.~\ref{sec:observations} introduces the observations and the reduction methods. In Sect.~\ref{sec:observations_analysis} we describe the observations in terms of their signal-to-noise ratio (S/N) and its relation to interstellar scintillation. In Sect.~\ref{sec:timing_analysis} we present the timing results and we study the influence of the S/N and bandwidth (BW) on the timing analysis; further details on this analysis are provided in the Appendix~\ref{sec:appendices}. In Sect.~\ref{sec:enterprise} we use the \texttt{ENTERPRISE} software to perform a noise analysis of the observations and estimate the contribution of a GWB. Finally, in Sect.~\ref{sec:conclusions} we present the main conclusions of our study. 

%
\section{Observations \label{sec:observations}}
%

As described in more detail in \cite{Gancio2020}, the design of the antennas allows us to observe a source continuously over 220 minutes. Their receivers are not currently refrigerated and have a system temperature of $T_\mathrm{sys} \sim 100$~K. The back-end is based on two software-defined radios 
which acquire raw samples with a maximum rate of 56~MHz per board. A1 uses these two digital plates in consecutive radio frequencies with a total bandwidth of $112$~MHz in a single polarization mode, while A2 uses those digital plates in one per polarization, thus covering a bandwidth of $56$~MHz. Those characteristics are summarized in Table \ref{table:par_obs}. 
In 2019 November, A1's receiver front-end went into commissioning. The electronics and systems were verified and improved, resulting in a slightly higher sensitivity and the recovery of the second polarization. Nonetheless, the observations were retaken with the previous configuration to have a homogeneous data set.

\begin{table}[h]
	{\centering
	\caption{Parameters of the Observations} \label{table:par_obs}
	\begin{tabular}{l c c}
		\hline \hline
		 & A1 & A2 \\
		\hline
		\!Number of observations & 170 (145*) & 197 (171*) \\
		\!MJD start -- MJD finish & \multicolumn{2}{c}{58596.7 -- 58999.6} \\
		\!Total observation time [h] & 391 (372*) & 393 (381*) \\
		\!Central frequency [MHz] & 1400, 1415, 1428 & 1428  \\
		\!Bandwidth ($BW$) & 112 MHz & 56 MHz \\
		\!Polarization modes & 1 & 2  \\
		\!Frequency channels ($n_\mathrm{chan}$) & 64/128 & 64 \\
		\!Time resolution  [$\mu$s] & \multicolumn{2}{c}{73.14}\\
		\!Phase bins ($n_\mathrm{bin}$) & \multicolumn{2}{c}{512/1024} \\
		\hline \hline
	\end{tabular}
	\par}
	\textbf{Note.} Values marked with (*) correspond to the restricted data set used in Sec.~\ref{sec:enterprise} (observations lasting more than 40 min that achieve a $\mathrm{S/N}>40$ and $\sigma_\mathrm{TOA}<1~\mu$s).
\end{table}

In this work we present the analysis of a data set of 170 observations with A1 and 197 with A2 over an interval of 13 months, from 2019 April 23 to 2020 May 30. This includes days with multiple observations (89 days with two observations, 24 days with three, and one day with four). 
The observations add up to over 390 hs of observation with each antenna (Table~\ref{table:par_obs}), leading to an observation efficacy of 
0.26 for both antennas. This efficacy is aimed to be improved in a future considering that (i) A1 underwent maintenance between 2019 October 8 and November 29, (ii) an unusually loud source of local radio-frequency interference (RFI) was particularly active in 2019 June-July during morning time, affecting A1 more notably, (iii) during 2020 February the observations stopped due to tests in the new automated pointing software and scheduler, (iv) A2 had lost observing time due to problems with a hard disk.

The receptor in A1 became more sensitive to local RFI after its upgrade in 2019 December. We found that the program \texttt{RFIClean}\footnote{\url{https://github.com/ymaan4/rfiClean}} (\cite{Maan:2020ent}) gave better results than the \texttt{rfifind} task in \texttt{PRESTO} to clean RFI. We therefore ran both software programs in all A1 observations carried out from 2019 November onwards.

The observations, stored in \textit{filterbank} format,\footnote{\url{http://sigproc.sourceforge.net/}} were folded and de-dispersed with \texttt{PRESTO} \citep{Ransom2003, Ransom2011} using $n_\mathrm{bins}=512$ or $1024$ phase bins\footnote{In Appendix~\ref{sec:timingnbins} we show that the number of phase bins does not affect the posterior analysis as long as $n_\mathrm{bins} \geq 256$.} and $n_\mathrm{chan}=64$ frequency channels for A2 observations and $n_\mathrm{chan}=64$ or $128$ for A1 observations. The data were folded using the \texttt{timing} flag of the task \texttt{prepfold} and the parameter (\texttt{.par}) file provided by IPTA,\footnote{\url{http://ipta4gw.org//data-release/}} ``Combination B'' with edits adapted to the IAR site. 
We then calculated the time of arrival (TOA) of the pulses using the \texttt{pat} package in \texttt{PSRCHIVE} \citep{Hotan:2004tz} with a Fourier phase gradient-matching template fitting \citep{Taylor1992}. The template was obtained applying a smoothing wavelet algorithm to a best profile; a more detailed discussion of the template selection is provided in Appendix~\ref{sec:templates}. The TOAs in this data set were fixed of clock systematics on 2019 April 22 (MJD 58595), when we reached an accuracy of $< 1~\mu$s \citep[see][for details on clock settings]{Gancio2020}.

%
\section{Analysis of the observations \label{sec:observations_analysis}}
%

\subsection{S/N of the observations} \label{sec:SNobs}

\begin{figure}[h]
    \centering
    \includegraphics[width=\linewidth]{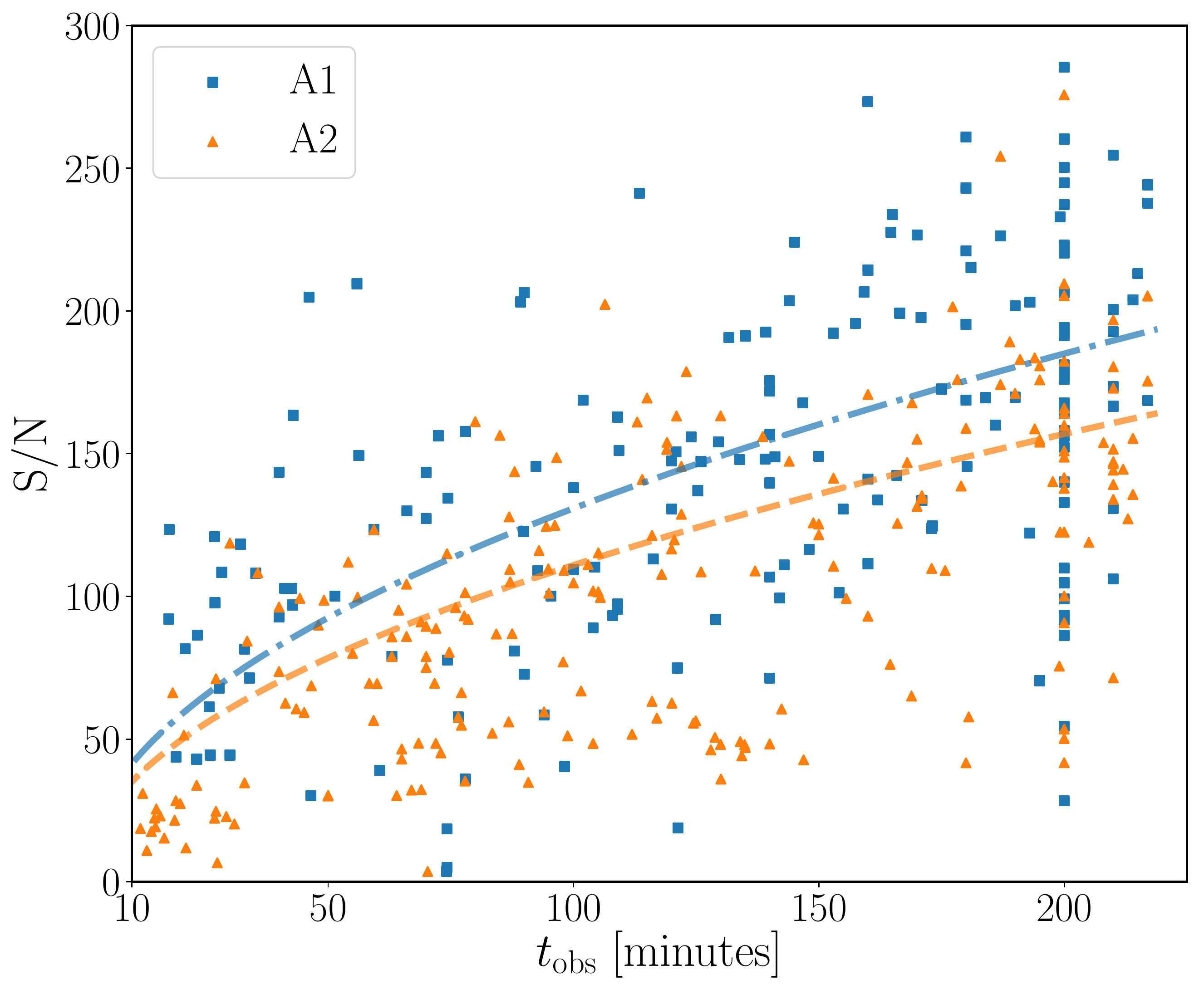}
    \caption{Signal-to-noise ratio of the observations of each antenna as a function of their $t_\mathrm{obs}$. We also plot $f(t_\mathrm{obs})=a \sqrt{t_\mathrm{obs}}$, where $a=13.1~\mathrm{min}^{-1/2}$ for A1 and $a=11.1 ~\mathrm{min}^{-1/2}$ for A2.} 
    \label{fig:tobs_sn}
\end{figure}

In order to characterize the S/N of the observations we use the functions \texttt{getDuration} and \texttt{getSN} of the Python package \texttt{PyPulse}\footnote{\url{https://github.com/mtlam/PyPulse}} \citep{PyPulse}. In Fig.~\ref{fig:tobs_sn} we show the S/N of each observation as a function of its duration. The mean S/N of observations with A1 is 151 and with A2 is 105, with mean observing times of 147 minutes and 116 minutes, respectively. However, we note that these numbers are affected by many short and low-quality observations.

When we restrict our analysis to observations with $\mathrm{S/N}>50$, the mean S/N for observations with A1 increases to 166 and with A2 to 122, with mean observing times of 162 minutes and 124 minutres, respectively. We summarize these and other values in Table~\ref{table:nobs_sn}. 

We observe a positive correlation between S/N and $t_\mathrm{obs}$, fitting to a $\mathrm{S/N} \propto \sqrt{t_\mathrm{obs}}$ as expected \citep{lorimer2012handbook}:
\begin{equation} \label{eq:SN}
   S/N 
   = \sqrt{n_\mathrm{P} \: t_\mathrm{obs} \: BW } \left( \frac{ T_\mathrm{peak} }{ T_\mathrm{sys}} \right) \frac{\sqrt{W(P-W)}}{P},
\end{equation}
\noindent where $P$ is the pulsar period and $W$ its width, $T_\mathrm{peak}$ is its maximum amplitude, $T_\mathrm{sys}$ is the noise temperature of the system, $t_\mathrm{obs}$ is the observing time,
and $n_\mathrm{P}$ is the number of polarizations observed.

\begin{figure}[h]
    \centering
    \includegraphics[width=\linewidth]{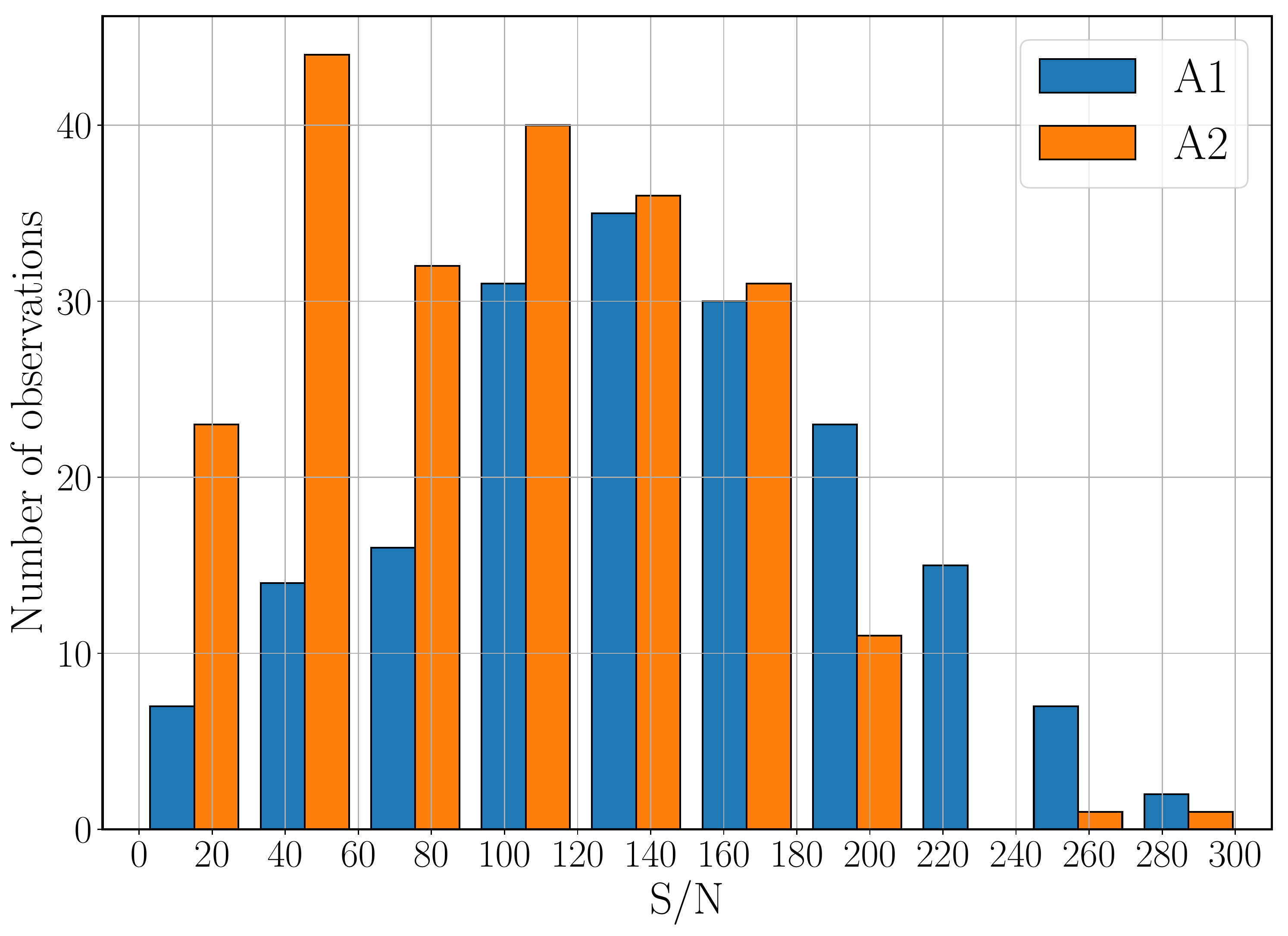}
    \caption{Histogram of the observations for each antenna, A1 and A2, according to their S/N.}
    \label{fig:histogramas}
\end{figure}

We collect the observations per S/N for each antenna and display them 
as histograms in Fig.~\ref{fig:histogramas}. 
We observe a distribution for A1 with a mean higher than the corresponding distribution for A2, perhaps due to the broader band sensitivity of A1.

We collect the observations into sets of $\mathrm{S/N}>1$, 50, 80, 110, 140, and 170, corresponding to roughly 
\textbf{}the position of the larger bins in the A2 histogram. In Table~\ref{table:nobs_sn} we specify the number of observations, mean duration, and mean S/N for each of these sets.

\setlength{\tabcolsep}{3.5pt}
\begin{table*}[htb] 
    \centering
    \begin{tabular}{l | ccc | ccc | ccc | ccc | ccc | ccc }
    \hline\hline    
    & \multicolumn{3}{c|}{S/N $>1$} & \multicolumn{3}{c|}{S/N $>50$} & \multicolumn{3}{c|}{S/N $>80$} & \multicolumn{3}{c|}{S/N $>110$} & \multicolumn{3}{c|}{S/N $>140$} & \multicolumn{3}{c}{S/N $>170$} \\
     & $N$ & $\langle \mathrm{S/N} \rangle$ & $\langle t_\mathrm{obs} \rangle$ & $N$ & $\langle \mathrm{S/N} \rangle$ & $\langle t_\mathrm{obs} \rangle$ & $N$ & $\langle \mathrm{S/N} \rangle$ & $\langle t_\mathrm{obs} \rangle$ & $N$ & $\langle \mathrm{S/N} \rangle$ & $\langle t_\mathrm{obs} \rangle$ & $N$ & $\langle \mathrm{S/N} \rangle$ & $\langle t_\mathrm{obs} \rangle$ & $N$ & $\langle \mathrm{S/N} \rangle$ & $\langle t_\mathrm{obs} \rangle$\\
    \hline
    A1 & 170 & 151 & 147 & 159 & 160 & 155 & 150 & 166 & 160 & 120 & 183 & 166 & 96 & 197 & 180 & 59 & 223 & 187 \\
    A2 & 197 & 105 & 116 & 164 & 120 & 121 & 128 & 136 & 146 & 88 & 153 & 178 & 58 & 168 & 192 & 22 & 192 & 194 \\
    A1+A2 & 367 & 127 & 130 & 323 & 140 & 140 & 278 & 152 & 154 & 208 & 170 & 171 & 154 & 186 & 182 & 81 & 214 & 191 \\
    \hline
    \end{tabular}
    \caption{Number of observations $N$, mean S/N, and mean $t_\mathrm{obs}$ expressed in minutes per S/N subset per antenna.}
    \label{table:nobs_sn}
\end{table*}

\subsection{Scintillations}

In what follows we assume that the expected S/N scales $\propto \sqrt{t_\mathrm{obs}}$ and that additional variations in the S/N are due to scintillation. We note that the observations described in Sec.~\ref{sec:observations} lack of absolute flux calibrations and thus possible variations in $T_\mathrm{sys}$ are not accounted for. Moreover, RFI is also variable and its mitigation leads to variations in the effective bandwidth of each observation, so additional dispersion in the S/N vs. $t_\mathrm{obs}$ relation is also expected. 
\begin{figure*}[htb]
    \centering
    \includegraphics[width=0.496\linewidth]{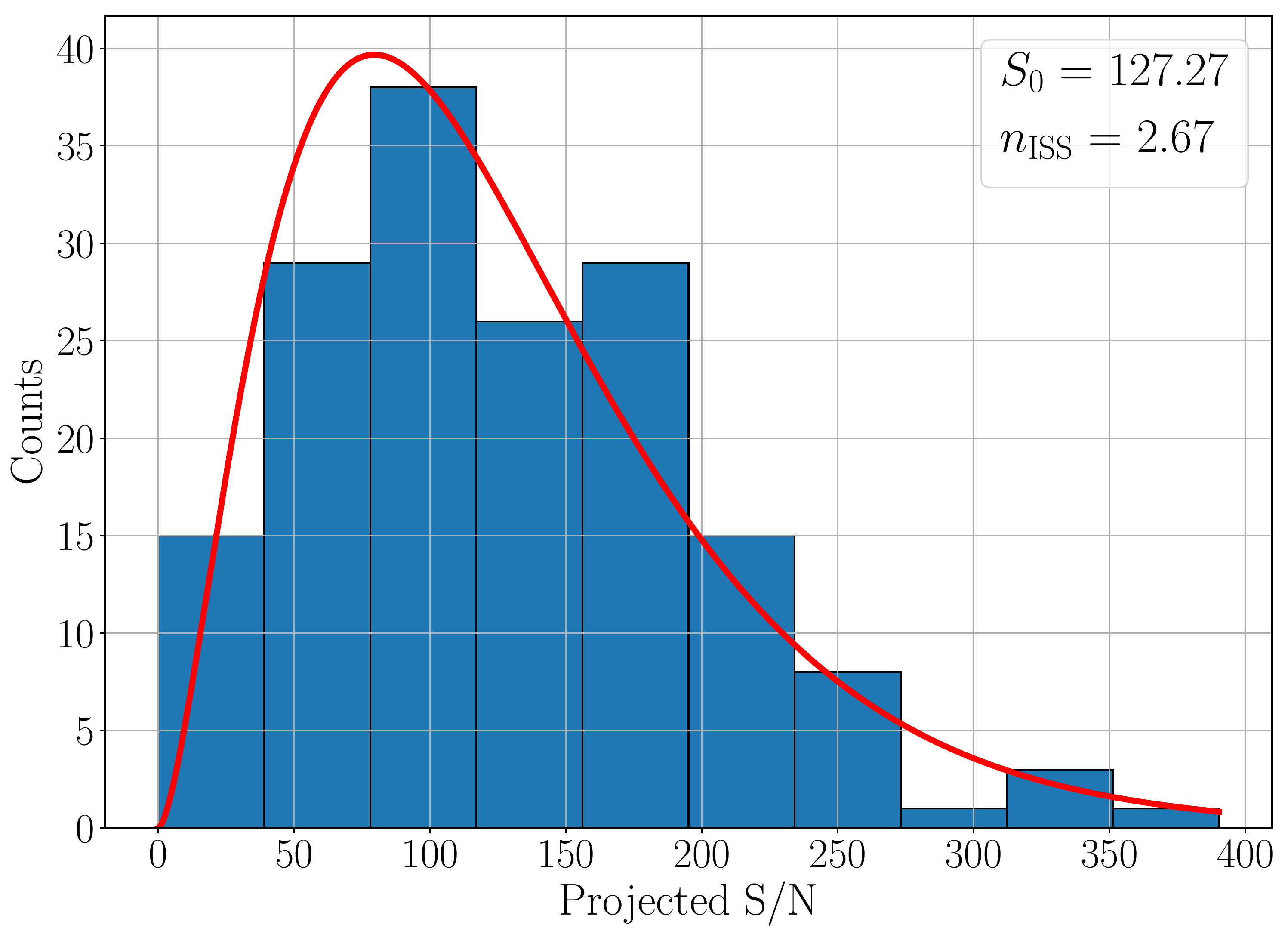}
    \includegraphics[width=0.496\linewidth]{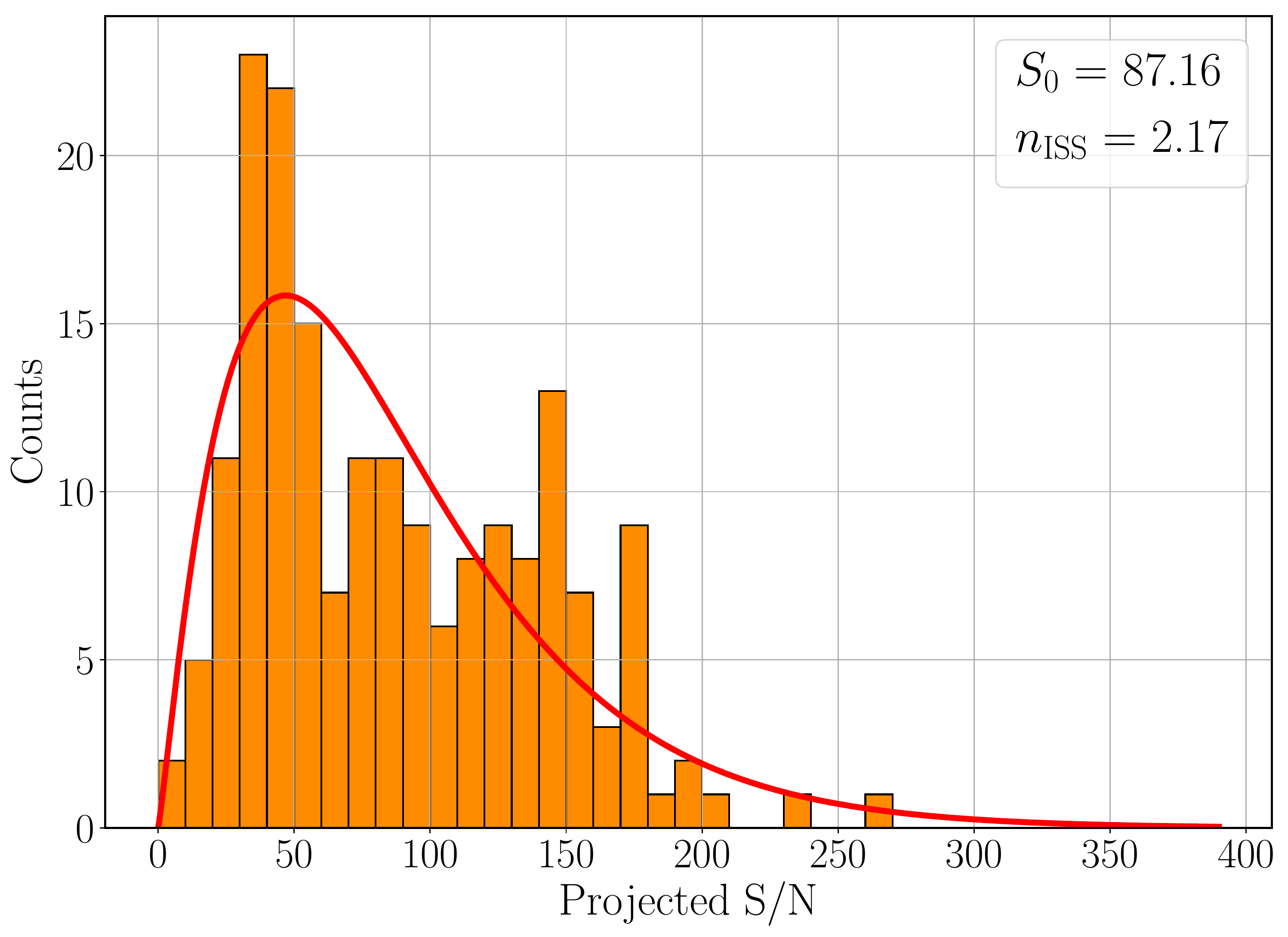} \\
    \includegraphics[width=0.496\linewidth]{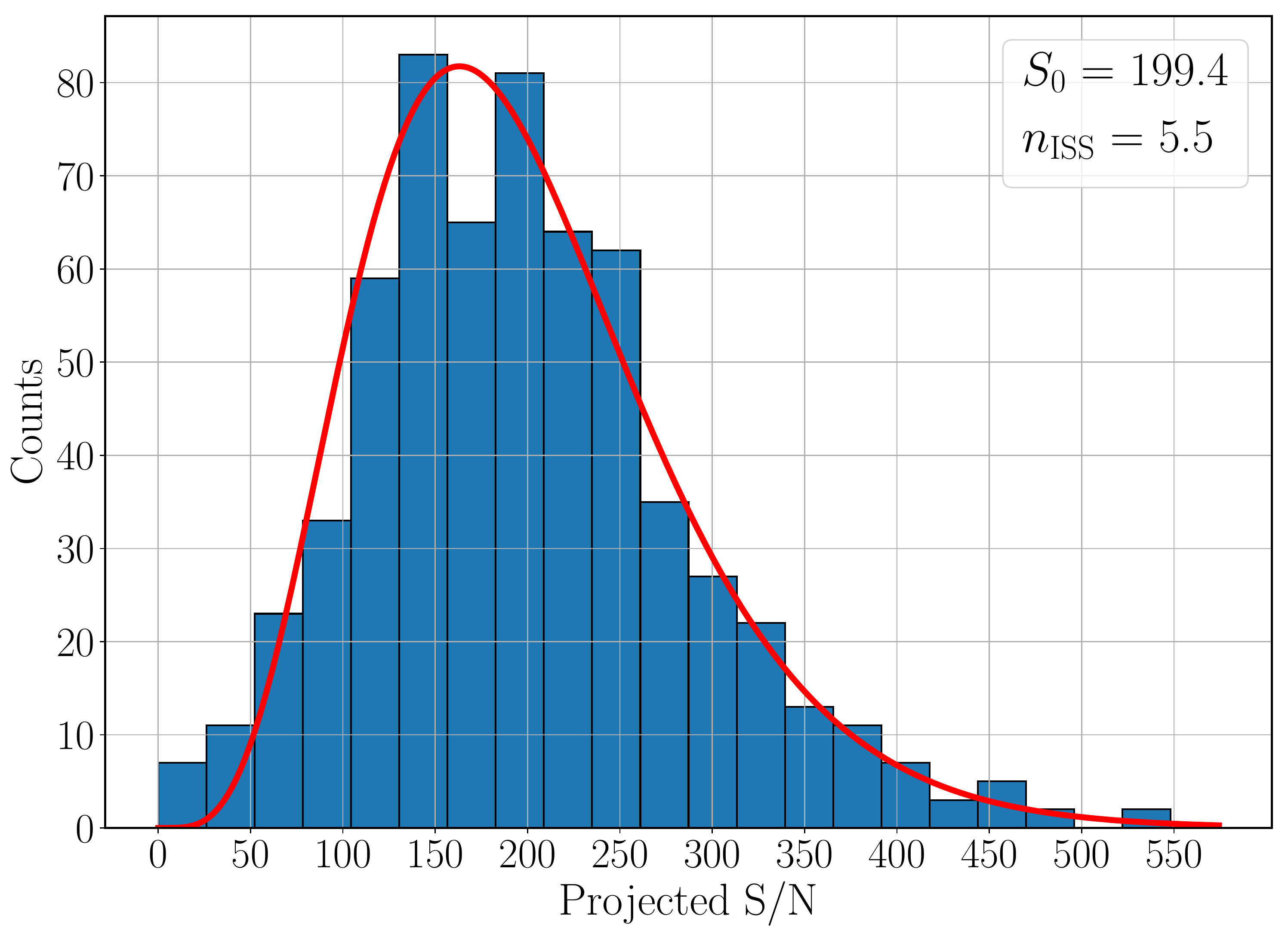}
    \includegraphics[width=0.496\linewidth]{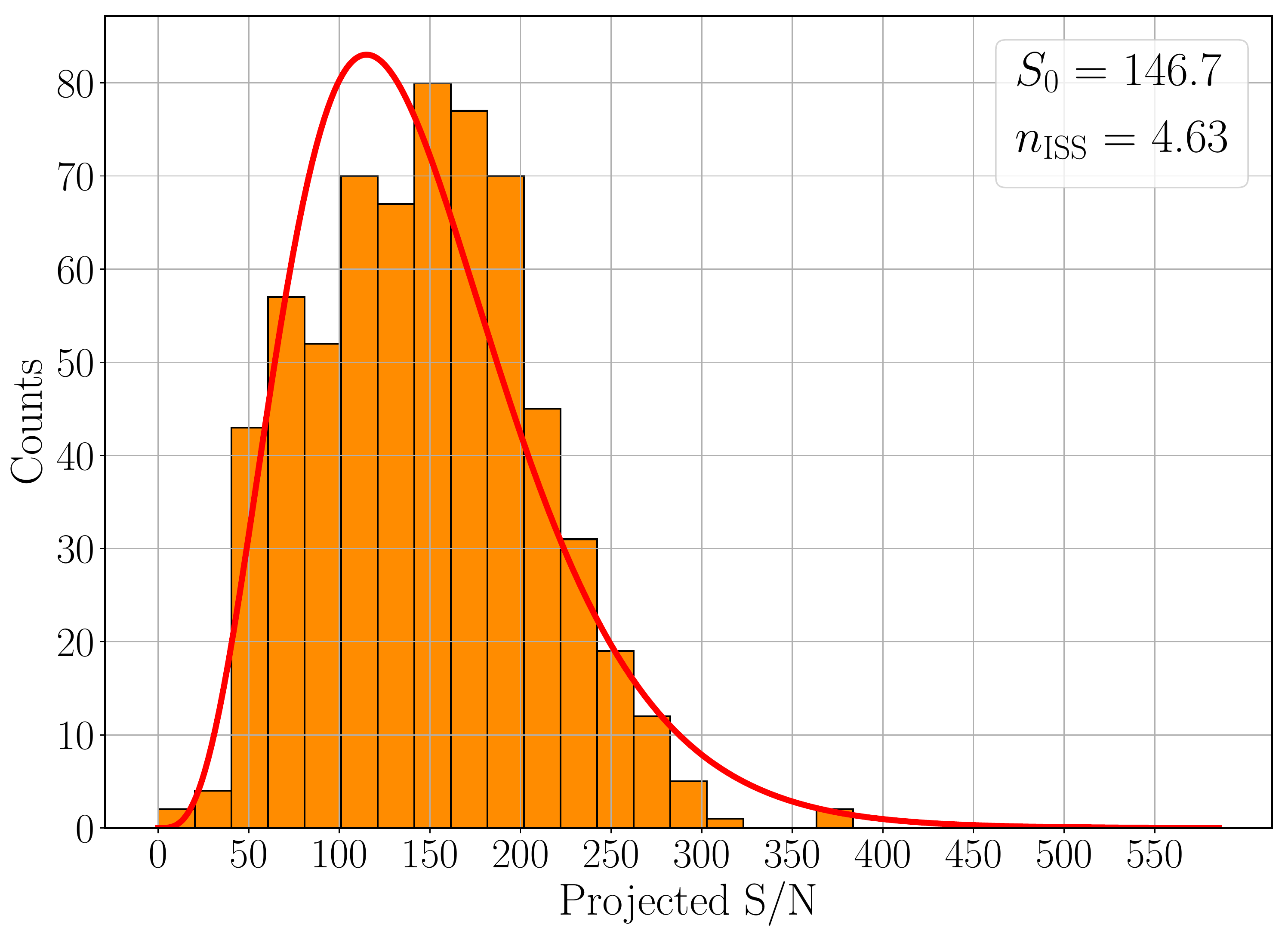}
    \caption{Histograms of projected pulse S/N for J0437$-$4715 for A1 (left column) and A2 (right column). The top row is for the full observations and the bottom row for observations split in segments such that $t_\mathrm{min}=2000$~s.
    The line shows the estimated scintillation distribution from fitting $n_\mathrm{ISS}$ in Eq.~\ref{eq:scintillation}. } 
    \label{fig:scintillation}
\end{figure*}

To quantify the variations due to scintillation we build a projected pulse S/N as $\mathrm{S/N}_\mathrm{proj} = \mathrm{S/N} \sqrt{t_\mathrm{max}/t_\mathrm{obs}}$, with $t_\mathrm{max} = 217$~min. Given that short observations have a large uncertainty in their determined S/N, we only use observations with $t_\mathrm{obs} > 20$ minutes (which is roughly half of the scintillation timescale).
Fig.~\ref{fig:scintillation} shows a histogram of the projected pulse S/N for A1 and A2. The line shows the estimated probability density function (PDF) from scintillation \citep{Cordes:1997my}
\begin{equation}
    f_S(S|n_\mathrm{ISS}) = \frac{\left(S n_\mathrm{ISS}/S_0 \right)^{n_\mathrm{ISS}}}{S\Gamma(n_\mathrm{ISS})}\exp{\left( \frac{-S n_\mathrm{ISS}}{S_0} \right)} \; \Theta(S),
    \label{eq:scintillation}
\end{equation} 
where $n_\mathrm{ISS}$ is the number of scintles, $S_0$ is the mean value of the signal $S$ (i.e., $S_0 = \langle \mathrm{S/N} \rangle$), and $\Theta$ is the Heaviside step function. We calculate $n_\mathrm{ISS}$ by fitting the normalized\footnote{We normalize the number of observations in each S/N bin by the total number of observations of each antenna.} data for each antenna. We obtain $n_\mathrm{ISS}=2.67 \pm 0.31$ for A1 and $n_\mathrm{ISS}=2.17 \pm 0.25$ for A2, with $S_{0} = 127.27$ for A1 and $S_{0}=87.16$ for A2. The bin size is determined using the Knuth's rule \citep{Knuth2006} algorithm provided in \texttt{astropy} \citep{astropy2013,astropy2018}, though we confirm that the obtained values do not depend on the binning by repeating the analysis for different bin sizes. 

In addition, we make use of the long duration of the observations, which is significantly larger than the typical scintillation timescale for J0437$-$4715. We split the observations in segments lasting $t_\mathrm{min}=2000$~s and $t_\mathrm{min}=5000$~s and repeat the previous analysis. In this case we obtain larger values of $n_\mathrm{ISS} \sim 5$. 

For each of these fittings we perform a Kolmogorov-Smirnov (KS) test for goodness of fit. This test quantifies the distance between the empirical distribution of the sample (obtained from the projected S/N) and the cumulative distribution function of the reference distribution (obtained from fitting $n_\mathrm{ISS}$ in Eq.~\ref{eq:scintillation}) under the null hypothesis that the sample is drawn from the reference distribution. 
The null hypothesis can be rejected at a given confidence level $\alpha$ if the resulting $p$-value is lower than $1-\alpha$. The $p$-values obtained are summarized in Table~\ref{table:niss_table}. For $\alpha=0.9$ ($90\%$ confidence level) we find that the goodness of fit cannot be statistically rejected for complete observations with either A1 or A2, or split observations of A1, all of which have a large $p$-value. However, the fits to the split observations of A2 fail this test, suggesting that, for short observations with A2, Eq.~\ref{eq:scintillation} may not be entirely valid or that the estimate of the projected S/N becomes unreliable. 

\begin{table}[h]
    \centering
    \caption{Adjusted values of $n_\mathrm{ISS}$ for each set of observations and the Kolmogorov-Smirnov test $p$-value for each fitting.} 	\label{table:niss_table}
    \begin{tabular}{c|ccc|ccc}
    \hline \hline
    & \multicolumn{3}{c|}{A1} & \multicolumn{3}{c}{A2} \\
    & $n_\mathrm{ISS}$ & error & $p$ & $n_\mathrm{ISS}$ & error & $p$ \\
    \hline
    No split & 2.67 & 0.31 & 0.38 & 2.17 & 0.25 & 0.24 \\
    Split $t_\mathrm{min}=5000$~s &  6.33 & 0.54 & 0.90 & 5.53 & 1.04 & 0.009 \\
    Split $t_\mathrm{min}=2000$~s & 5.50 & 0.36 & 0.70 & 4.63 & 0.43 & 0.004 \\
    \hline
    \end{tabular}
\end{table}

We compare our values of $n_\mathrm{ISS}$ with theoretical estimations following \cite{Lam2020}. We scale the scintillation parameters given at the frequency of 1.5~GHz by \cite{Keith2013} to match our observations centered at 1.4~GHz and obtain the scintillation bandwidth $\Delta \nu_\mathrm{d} = 740$~MHz and scintillation timescale $\Delta t_\mathrm{d} = 2290$~s. We calculate $n_\mathrm{ISS}$ via the usual formula
\begin{equation}
    n_\mathrm{ISS} \approx \left(1+\eta_t\frac{T}{\Delta t_\mathrm{d}}\right)\left(1+\eta_\nu\frac{BW}{\Delta \nu_\mathrm{d}}\right)
\end{equation}
where $\eta_t$ and $\eta_\nu$ are filling factors $\sim 0.2$. The estimated $n_\mathrm{ISS}$ for $T=220$ minutes are 2.22 for A1 (BW~=~112~MHz) and 2.18 for A2 (BW~=~56~MHz). We confirm that the value obtained with A2 is consistent with the expectations, although for A1 it is larger than expected, perhaps due to additional factors affecting the variability observed.

\cite{2006A&A...453..595G} found two scintillation scales observing J0437$-$4715 in 327~MHz. Rescaling those scales to our observing frequency, 1400~MHz, we find time scales of $\Delta t_\mathrm{d,1} = 5727$~s and $\Delta t_\mathrm{d,2} = 515$~s, leading to $n_\mathrm{ISS,1}=1.46$ for both antennas 
and $n_\mathrm{ISS,2}=6.58$ for A1 and $n_\mathrm{ISS,2}=6.35$ for A2. 
The latter values are close to theose displayed in Table~\ref{table:niss_table} for the split observations, consistent with 
the shorter observations being more sensitive to the shorter-scale scintillations. Note also that those scintillations scales have been observed to vary notably between epochs \citep{Smirnova:2006mz}.

%
\section{Timing analysis \label{sec:timing_analysis}}
%

Here we discuss the timing-error dependence on three parameters: (i) the S/N of the observations, (ii) the number of bins used in the reduction of the observations, and (iii) the BW of the observations. In addition we study and quantify other sources of systematic errors.

\subsection{Timing Residuals} \label{sec:timing_residuals}

We compute the timing residuals of the TOAs using \texttt{Tempo2} \citep{Hobbs2006} and its Python wrapper, \texttt{libstempo} \citep{Vallis2020}, with the timing model given in the file \texttt{J0437$-$4715.par} provided by IPTA and adapted to the IAR observatory\footnote{In this \texttt{.par} we also included four \texttt{JUMPs}  to account for the different central frequencies of the observations, and the corresponding antenna (A1/A2; see Table~\ref{table:par_obs}).}.
\texttt{Tempo2} returns: (i) the MJD, residual, and template-fitting error ($\sigma_\mathrm{TOA}$) of each observation, and (ii) the timing model parameters, the weighted errors of the residuals (rms), and the $\chi^2_\mathrm{red} = \chi^2/n_\mathrm{free}$ of the timing model fit to the residuals. The $\chi^2$ test considers a good fit when $\chi^2_\mathrm{red} \sim 1$; instead, a value of $\chi^2_\mathrm{red} \gg 1$--assuming the timing model is correct--indicates the presence of outliers or an underestimation of the residuals errors. In this case we can: 
\begin{enumerate}
    \item assume a certain systematic error in the computation of the TOAs due, for instance, to instrumental errors such as observation \textit{timestamp}, reduced BW, hidden RFI, etc.;
    \item define a criterion to discard the outliers, for instance by vetting residuals above a certain value.
\end{enumerate}

\begin{figure}[h]
    \centering
    \includegraphics[width=\linewidth]{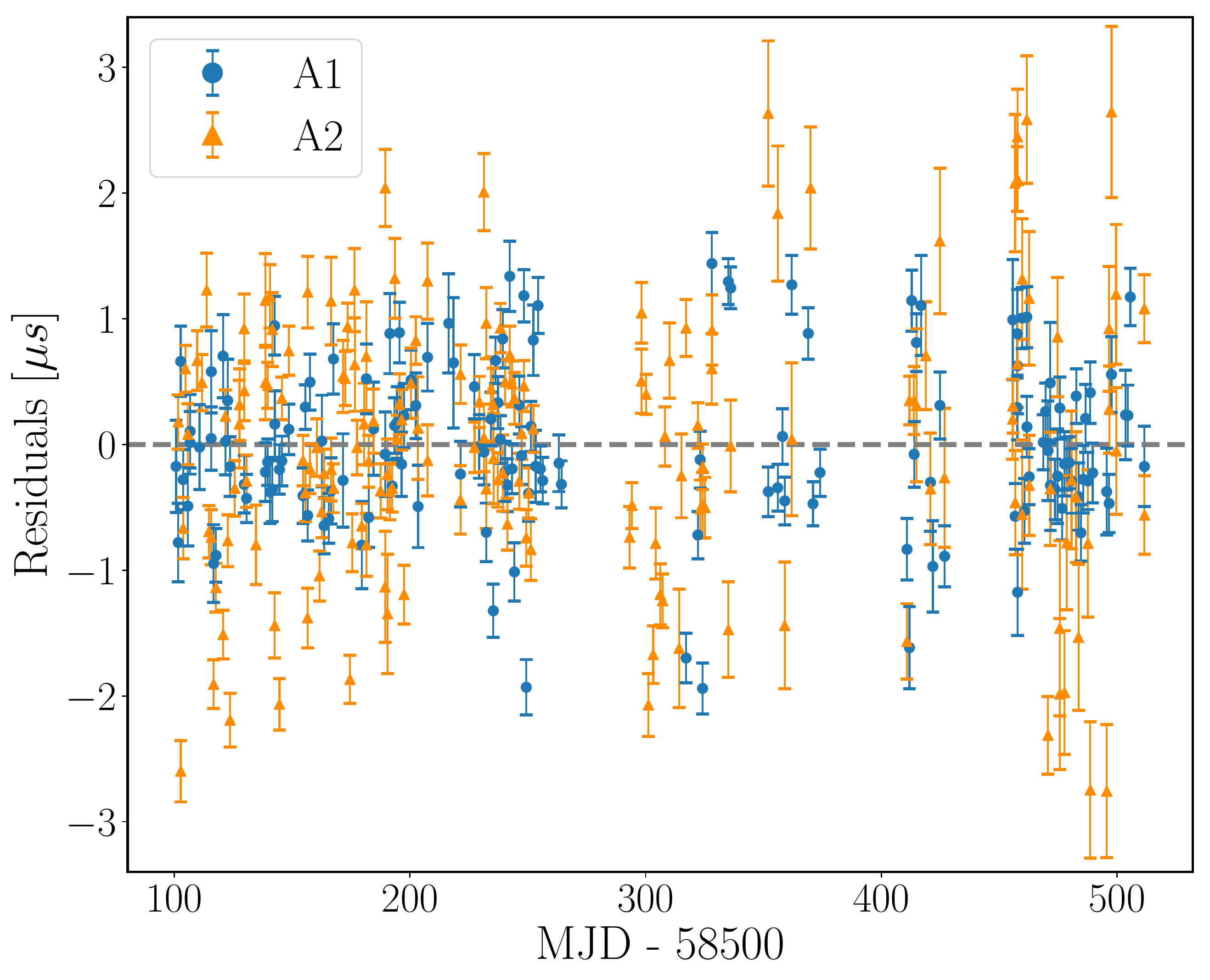}
    \caption{Timing residuals for the complete data set for A1 and A2.}
    \label{fig:residues_total}
\end{figure}
Fig.~\ref{fig:residues_total} shows the timing residuals of the observations taken with each antenna. The values of the $\chi^2_\mathrm{red}$ from the fits are greater than $1$, indicating the presence of outliers or underestimated errors. 

To account for possible systematic errors, we adopt a simplified approach\footnote{
In Sec.~\ref{sec:enterprise} we compare the results of this simplified model with those obtained using a standard and a more refined white noise model as in \cite{Arzoumanian2016}. 
} in which we add quadratically a common $\sigma_\mathrm{sys}$ to all the $\sigma_\mathrm{TOA}$, producing a total error 
\begin{equation} \label{eq:sigma}
   \sigma_\mathrm{tot}^2 = \sigma_\mathrm{TOA}^2 + \sigma_\mathrm{sys}^2.
\end{equation}

We calculate the value of $\sigma_\mathrm{sys}$ that leads to $\chi^2_\mathrm{red}=1$, obtaining $\sigma_\mathrm{sys} \sim 0.67~\mu$s for the observations with A1 and $\sigma_\mathrm{sys} \sim 1.0~\mu$s for the observations with A2. We recompute the rms using the corrected errors in the residuals by adding $\sigma_\mathrm{sys}$ as in Eq.~\ref{eq:sigma}. We obtain $\mathrm{rms}= 0.72~\mu$s for A1 and $\mathrm{rms} = 1.05~\mu$s for A2. 

In order to determine the effect of the outliers measurements we set a 3$\sigma$ criterion, but since the $\sigma_\mathrm{tot}$ itself depends on the assumed value of $\sigma_\mathrm{sys}$, we apply the following iterative process.
\begin{enumerate}
    \setlength{\itemsep}{0.5pt}
    \setlength{\parskip}{0.5pt}
    \item Given an initial $\sigma_\mathrm{sys}^{(i)}$ (as obtained previously), to each TOA we assign an error ${\sigma_\mathrm{tot}^{(i)\:2}} = \sigma_\mathrm{TOA}^{2} + {\sigma_\mathrm{sys}^{(i)\:2}}$.
    \item If the residual of an observation is such that $\left| \delta t \right| > 3 \sigma_\mathrm{tot}^{(i)}$, then this observation is discarded as an outlier.
    \item If the residual is such that $\left| \delta t \right| \leq 3 \sigma_\mathrm{tot}^{(i)}$, then we keep this observation and its TOA error is given the new value
    \begin{equation}
        {\sigma_\mathrm{tot}^\mathrm{(i+1)}}^2 = 
        {\sigma_\mathrm{TOA}}^2 + {\sigma_\mathrm{sys}^{(i+1)}}^2,
    \end{equation}
    where $\sigma_\mathrm{sys}^{(i+1)}$ is chosen such that when the new residuals are computed we get $\chi_\mathrm{red}^2=1$. In practice, the process converges after 1 or 2 iterations.
\end{enumerate}
\noindent In this way, we eliminate all the outliers in our data set (five observations for A1 and 24 for A2) and obtain refined values of the systematic errors $\sigma_\mathrm{sys} \sim 0.50~\mu$s for A1, $\sigma_\mathrm{sys} \sim 0.66~\mu$s for A2, and $\sigma_\mathrm{sys} \sim 0.59~\mu$s for A1+A2. 

\subsection{Timing versus S/N} \label{sec:timingSN}

We study the timing residuals for each S/N subset for each antenna; these are shown in Fig.~\ref{fig:residues_sn_A1_A2}. By filtering out the low-S/N observations, those with large residuals are eliminated. Thus we conclude that outliers tend to have low S/N; we note, however, that some low-S/N observations also have small residuals. 

We perform a timing analysis for A1, A2, and A1+A2. In all cases--even for large S/N values--we obtain  $\chi^2_\mathrm{red} \gg 1$. We interpret this as indicative of unaccounted-for systematic errors and we perform the procedure detailed in Sec.~\ref{sec:timing_residuals} to find the values of $\sigma_\mathrm{sys}$ that lead to $\chi^2_\mathrm{red} \approx 1$. Taking as a reference the case for $\mathrm{S/N}>50$, we obtain $\sigma_\mathrm{sys}=0.5~\mu$s for A1, $0.66~\mu$s for A2, and $0.59~\mu$s for A1+A2. We note that these values change if we do not remove the 3$\sigma$ outliers, leading to  $\sigma_\mathrm{sys}=0.67~\mu$s for A1, $0.99~\mu$s for A2, and $0.83~\mu$s for A1+A2. 

\begin{figure}[h]
    \centering
    \includegraphics[width=\linewidth]{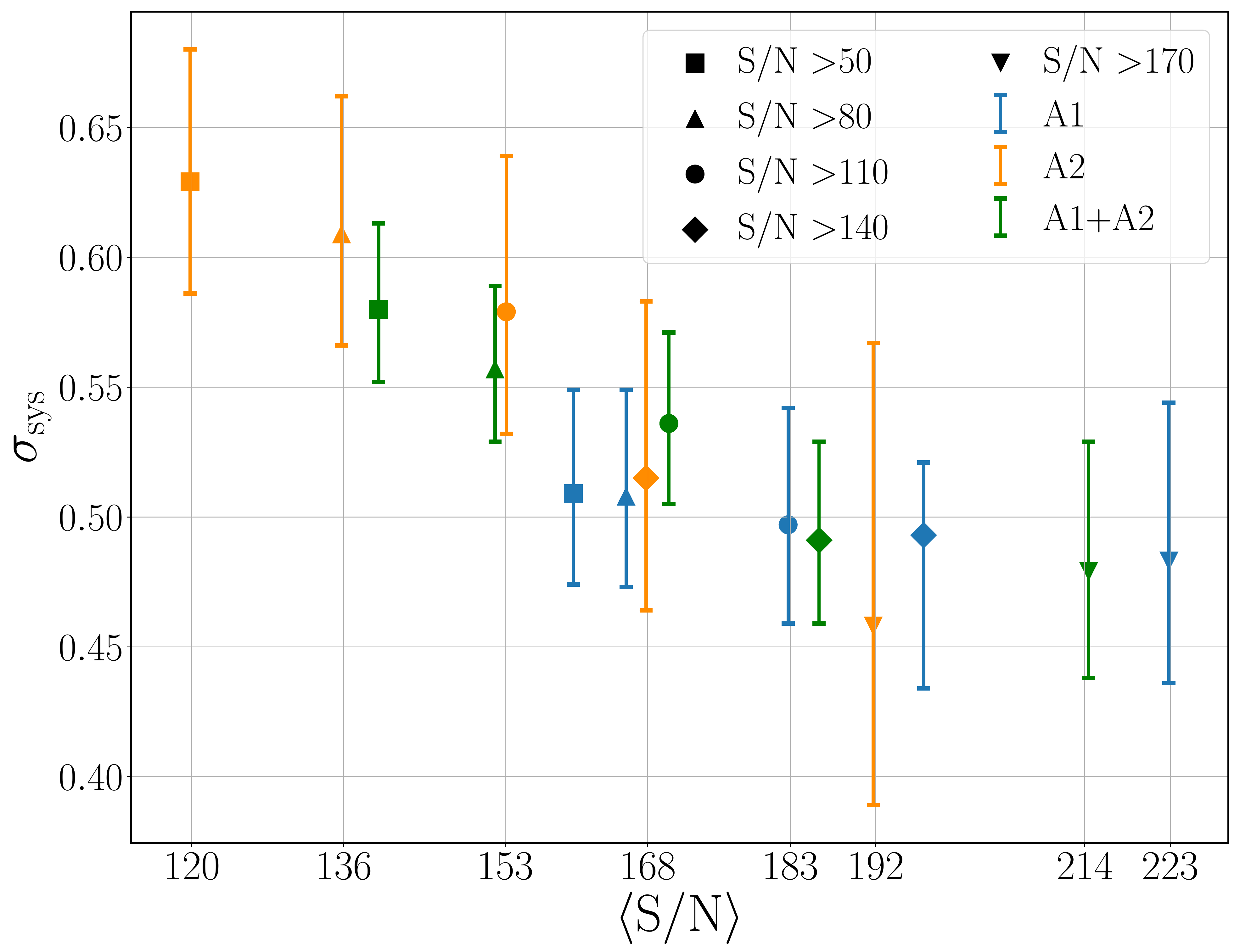}\\
    \includegraphics[width=\linewidth]{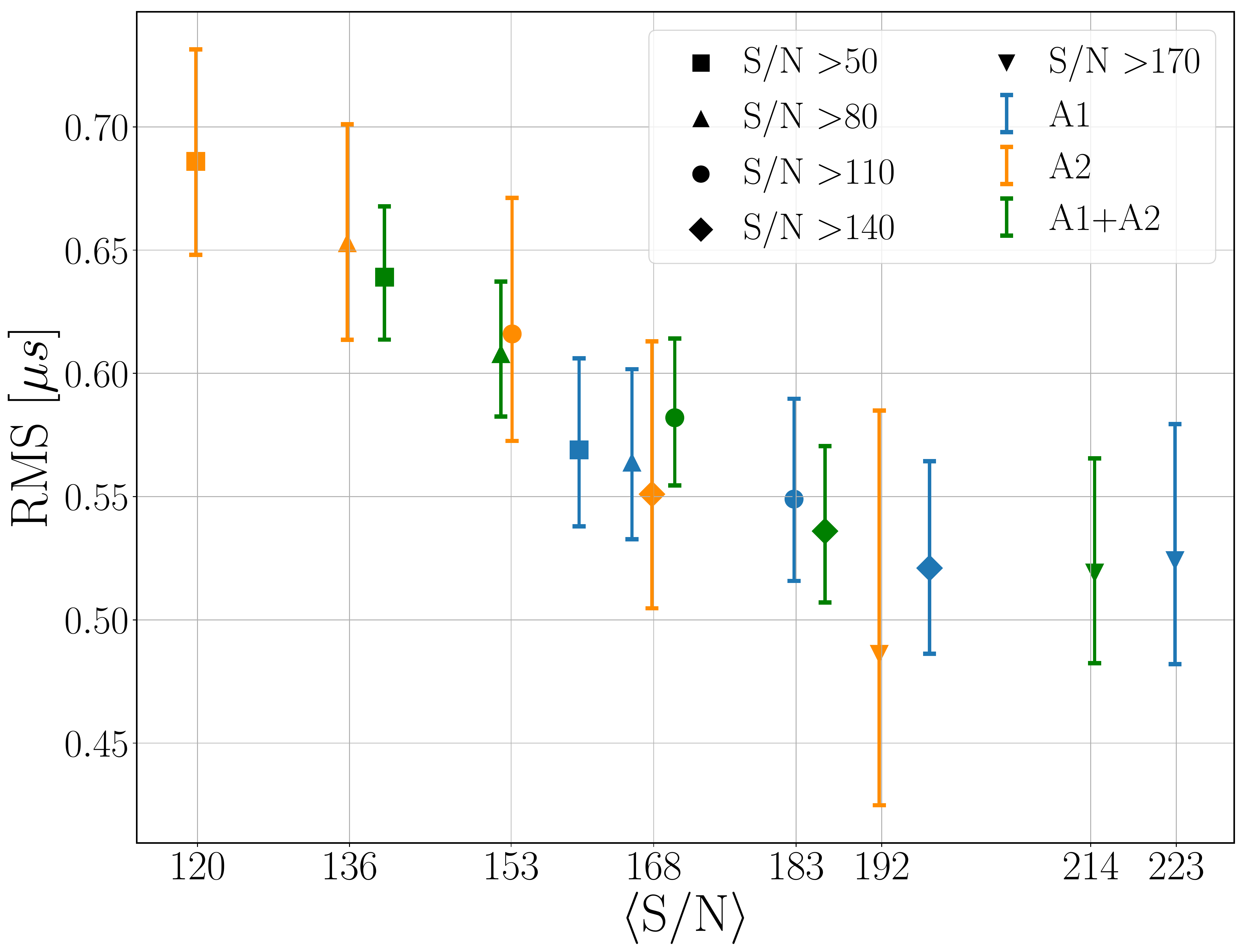}
    \caption{Top: $\sigma_\mathrm{sys}$ for the $ \langle \mathrm{S/N} \rangle $ of each subset of observations for each antenna and their corresponding error bars. Bottom: rms from recomputed time-of-arrival errors augmented by $\sigma_\mathrm{sys}$ and their corresponding error bars.}
    \label{fig:sn_sigma_rms}
\end{figure}

In Fig.~\ref{fig:sn_sigma_rms} we display the values of $\sigma_\mathrm{sys}$ and $\mathrm{rms}$ for each subset of observations with their 1$\sigma$ error bars ($\sim 68\%$ confidence limits). The error bars for $\sigma_\mathrm{sys}$ are computed as the values $\sigma_\mathrm{sys,min}$ that yield $\chi^2_\mathrm{red}(n_\mathrm{free},\alpha/2)$ and $\sigma_\mathrm{sys,max}$ that yield $\chi^2_\mathrm{red}(n_\mathrm{free},1-\alpha/2)$, with $\alpha=0.32$. 

The timing rms diminishes (i.e., improves) for higher S/N observations. The value of the rms is well-constrained to $\gtrsim 0.5~\mu$s, though values a little higher ($\approx 0.7~\mu$s) are obtained for A2 when low-S/N ($<100$) observations are included. In particular, for $\mathrm{S/N} >140$ we get rms $\approx 0.52~\mu$s for A1 and $0.55~\mu$s for A2, which is a slight improvement over those reported in \cite{Gancio2020} ($0.55~\mu$s for A1 and $0.81~\mu$s for A2). 

We also obtain a consistent value of $\sigma_\mathrm{sys} \approx 0.5~\mu$s. There is a systematic trend of lower $\sigma_\mathrm{sys}$ toward increasing S/N (Fig.~\ref{fig:sn_sigma_rms}), though with a small significance (close to or below 1$\sigma$ level). We conclude that the systematic errors of both IAR's antennas are of the order of $0.4$--$0.6~\mu$s when accounting for outliers and S/N effects.

Finally, the values of $\sigma_\mathrm{sys}$ and $\mathrm{rms}$ are smaller for A1 than for A2 for each subset of $\mathrm{S/N_{min}}$; this behavior subsists at the same $\langle \mathrm{S/N} \rangle$ (Fig.~\ref{fig:sn_sigma_rms}). 
Given that the main differences between the two antennas are BW and $n_\mathrm{P}$, we explore those dependences in detail to understand the reason(s) behind the improved timing precision of A1.
%

\subsection{Timing versus bandwidth} \label{sec:timing_bw}

In Sec.~\ref{sec:timingSN} we found that the A1 observations have a lower timing rms than the corresponding observations from A2. Since both antennas differ in their BW ($112$~MHz for A1 and $56$~MHz for A2), and $n_\mathrm{P}$ 
(1 and 2 for A1 and A2, respectively), we reduce the observations to the same BW and $n_\mathrm{P}$ in order to quantify the effect of those hardware differences on errors. For this analysis we use the six subsets of observations defined by their S/N in Sec.~\ref{sec:SNobs}. We split the A1 observations into two subintervals of BW$=56$~MHz using \texttt{pat} for the scrunching with the options \texttt{-j "F \{n\}"}, where $n=2$ is the number of subintervals. For the observations with A2 we would like to split the two polarizations separately; however, this is not possible as these observations only store the sum of both polarization modes. From the radiometer equation \citep{lorimer2012handbook}
\begin{equation} \label{eq:radiometre1}
\sigma_\mathrm{sys} \propto \frac{T_\mathrm{sys}}{\sqrt{n_\mathrm{P} \, BW}},
\end{equation}
we see that the errors scale with $n_\mathrm{P}^{-1/2}$; hence, we multiply the errors in the A2 residuals by a factor $\sqrt{2}$ to simulate a case with $n_\mathrm{P}=1$ (assuming we are not strongly affected by the polarization of the source). 
 
In this way, for each subset of $\mathrm{S/N_{min}}$ we have five groups of observations: three from A1 (one with BW$=112$~MHz and two reduced to $56$~MHz), and two from A2 (with errors modeled to two and one polarization modes). We model $\mathrm{S/N}(\sigma_\mathrm{TOA})$ and then compute $\langle \mathrm{S/N} \rangle$ for each of these subset from the $\sigma_\mathrm{TOA}$ of their observations.

\begin{figure}[h]
    \centering
    \includegraphics[width=\linewidth]{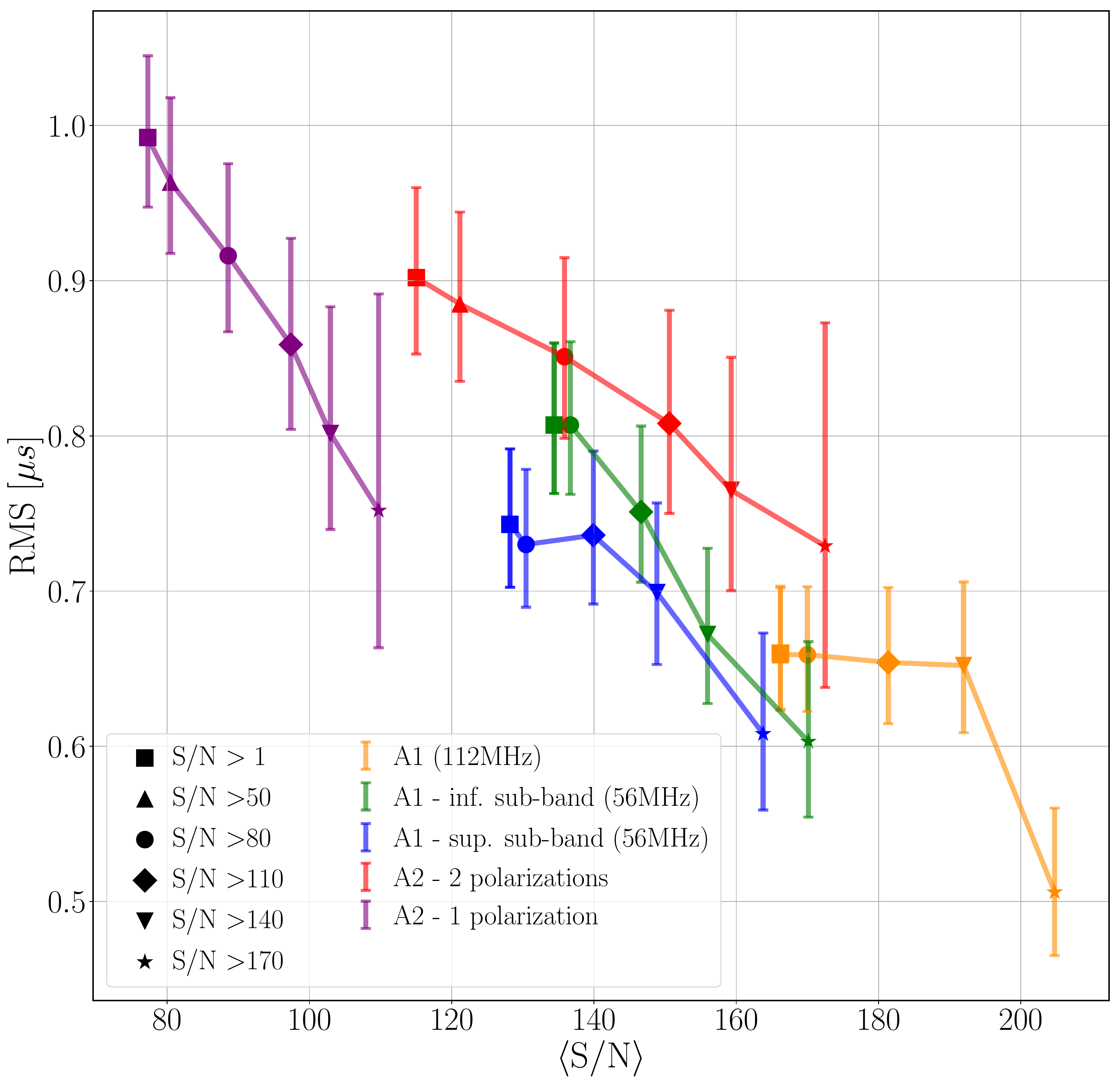}
    \caption{Timing residual rms values with the A1 observations scrunched to BW$=56$~MHz and those observed with A2 reduced to one polarization. We also reproduce the full A1 and A2 original residuals.}
    \label{fig:rms_bw}
\end{figure}

The resulting rms values are plotted in Fig.~\ref{fig:rms_bw}. For a given value of $\mathrm{S/N_{min}}$, the higher-frequency sub-band of the A1 observations has lower rms than the lower-frequency sub-band, which in turn is similar to the case of the A2 observations in one polarization. The inclusion of all the BWs for A1 or both polarizations of A2 shows consistently lower rms. These results can be interpreted as due to:
\begin{enumerate}
    \itemsep0em
    \item RFI affecting more the lower-frequency sub-band,
    \item effects of differential scintillation,
    \item dispersion effects being better modeled at higher frequencies.
\end{enumerate}

In conclusion, we found that the main difference in the timing errors between the antennas can be attributed to the difference in BW, followed by $n_\mathrm{P}$ and an increase of S/N in the selection of the observations. An increase in BW seems paramount for improving the timing errors.

\subsection{Timing versus observation length (with split of observations) \label{sec:timingsplit}}

The rms values improve both with longer observation times and with a higher number of TOAs \citep{lorimer2012handbook, Wang:2015bsa}. Here we investigate whether it is possible to improve the overall timing by splitting the long ($>200$ minute) observations into multiple subintegrations, producing various TOAs from each observation. In this way, we obtain additional data points at the expense of lower timing precision in each of them.

\begin{figure}[h]
    \centering
    \includegraphics[width=\linewidth]{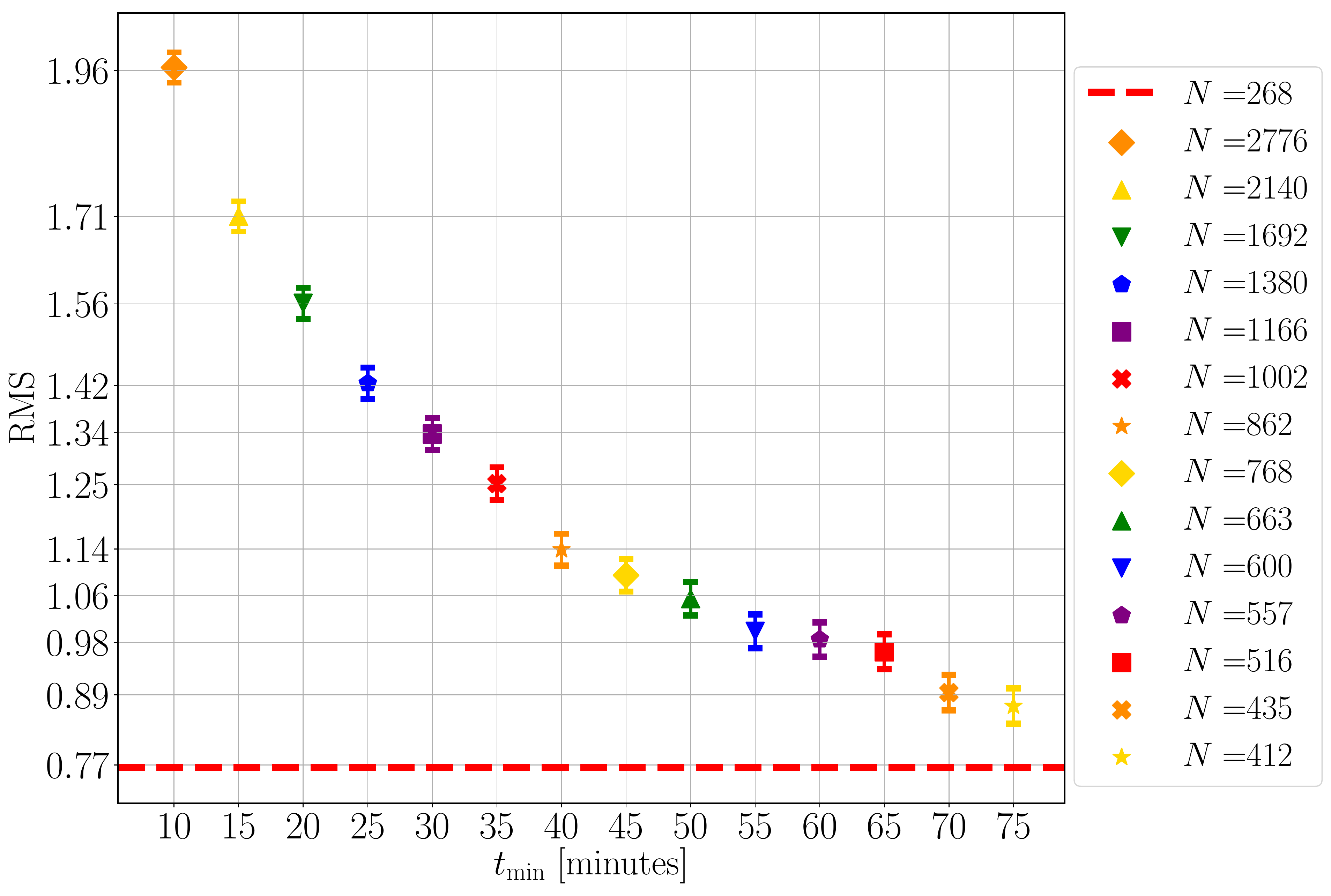}
    \caption{Timing obtained when splitting observations. The total number of points after splitting is detailed as $N$. The horizontal dashed line corresponds to no splitting. The projected crossing of curves occurs at about 90 minutes.}
    \label{fig:subints}
\end{figure}

We start with a set of 268 unsplit observations with $t_{\mathrm{obs}}>75$ minutes and $\sigma_{\mathrm{TOA}}<1.0~\mathrm{\mu s}$. First we calculate their rms (dashed line in Fig.~\ref{fig:subints}). Then we systematically split these observations for different values of the minimum duration of the subintervals considered, from 10 to 75 minutes (the shorter the subinterval, the larger the number of TOAs obtained). We plot the rms as a function of $t_\mathrm{min}$ in Fig.~\ref{fig:subints}, and specify the total number of points obtained from splitting in each case. We see that the rms diminishes monotonously as  $t_\mathrm{min}$ increases, showing that this method is not suitable for improving the timing of our observations. This is most likely a sign of the S/N being a major factor affecting our current timing precision; for $t_\mathrm{min} < 70$~min it is also possible that jitter affects the TOAs. 

%
%
\section{Noise analysis \label{sec:enterprise}}
%

In the following sections, we analyze: (i) the white noise in our data set, which is needed to estimate the systematic timing errors; (ii) the red noise, which is correlated in time and has a larger amplitude at low frequencies; (iii) the GWB at $\mu$Hz frequencies, which is produced from a variety of sources that we cannot identify individually. For this purpose, we use the software \texttt{ENTERPRISE} (Enhanced Numerical Toolbox Enabling a Robust PulsaR Inference SuitE), a pulsar-timing analysis code which performs noise analysis, GW searches, and timing-model analysis \citep{Ellis2019}. \texttt{ENTERPRISE} uses the timing model, previously fit with \texttt{Tempo2}, as the basis to construct a design matrix centered around the timing parameters. This is then used to find the maximum-likelihood fit for the white- and red-noise parameters.

\subsection{White-noise analysis \label{sec:white_noise}}

As described by \cite{Alam:2020fjy}, the white noise is modeled using three parameters.
\begin{enumerate}
    \item EQUAD accounts for sources of uncorrelated and systematic (Gaussian) white noise in addition to the template-fitting error in the TOA calculations.
    \item EFAC is a dimensionless constant multiplier to the TOA uncertainty from template-fitting errors.
    It accounts for possible systematics that lead to underestimated uncertainties in the TOAs.
    \item ECORR describes short-timescale noise processes that have no correlation between observing epoch, but are completely correlated between TOAs that were obtained simultaneously at different observing frequencies. This parameter accounts for wide-band noise processes such as pulse jitter \citep{Oslowski2011, Shannon2014}.
\end{enumerate}

Considering $\sigma_\mathrm{TOA}$ to the template-fitting error of a given observation, the resulting white-noise model is modeled by the noise covariance matrix \citep[][]{Lentati2014temponest}
\begin{multline}
    \sigma_{\nu \nu', tt'}^{2}= \delta_{tt'} [ \delta_{\nu \nu'} \left(\mathrm{EFAC}^{2}\, \sigma_\mathrm{TOA}^{2} + \mathrm{EQUAD}^2 \right) \\
    + \mathrm{ECORR}^{2} ],
\end{multline}
where $t$ and $\nu$ are the time and frequency of the observation, respectively.

Given that we have multiple TOAs per day (see Sec.~\ref{sec:observations}), we need to consider an ECORR contribution. We then incorporate all these noise components and timing-model parameters (as specified in Appx.~\ref{sec:reduction}) into a joint likelihood using \texttt{ENTERPRISE}. We sample the posterior distribution using the sampler \texttt{PTMCMCSampler} \citep{Ellis2017}, setting uniform prior distributions.

Firstly, we investigate the consistency between the analysis with \texttt{ENTERPRISE} and the independent analysis we presented in Sect.~\ref{sec:timingSN}. To this end, we use the same set of observations as in the aforementioned analysis while we fix the value $\mathrm{EFAC}=1$ and we exclude the ECORR parameter from our analysis, so that the Gaussian white noise EQUAD becomes equivalent to the parameter $\sigma_\mathrm{sys}$ in Eq.~\ref{eq:sigma}. We obtain a notorious agreement between the values of EQUAD and $\sigma_\mathrm{sys}$: when removing 3$\sigma$ outliers (Sect.~\ref{sec:timingSN}) we obtain $\mathrm{EQUAD} \approx 0.57~\mathrm{\mu}$s, fully consistent with the systematic error of $\sigma_\mathrm{sys} \approx 0.59~\mu$s that we found in Sect.~\ref{sec:timingSN} for A1+A2 and S/N$>50$; without removing the outliers, the results are $\mathrm{EQUAD} \approx 0.80~\mathrm{\mu}$s and $\sigma_\mathrm{sys} \approx 0.83~\mu$s, which again are fully consistent.

%

%

\begin{figure}[h]
    \centering
    \includegraphics[width=\linewidth]{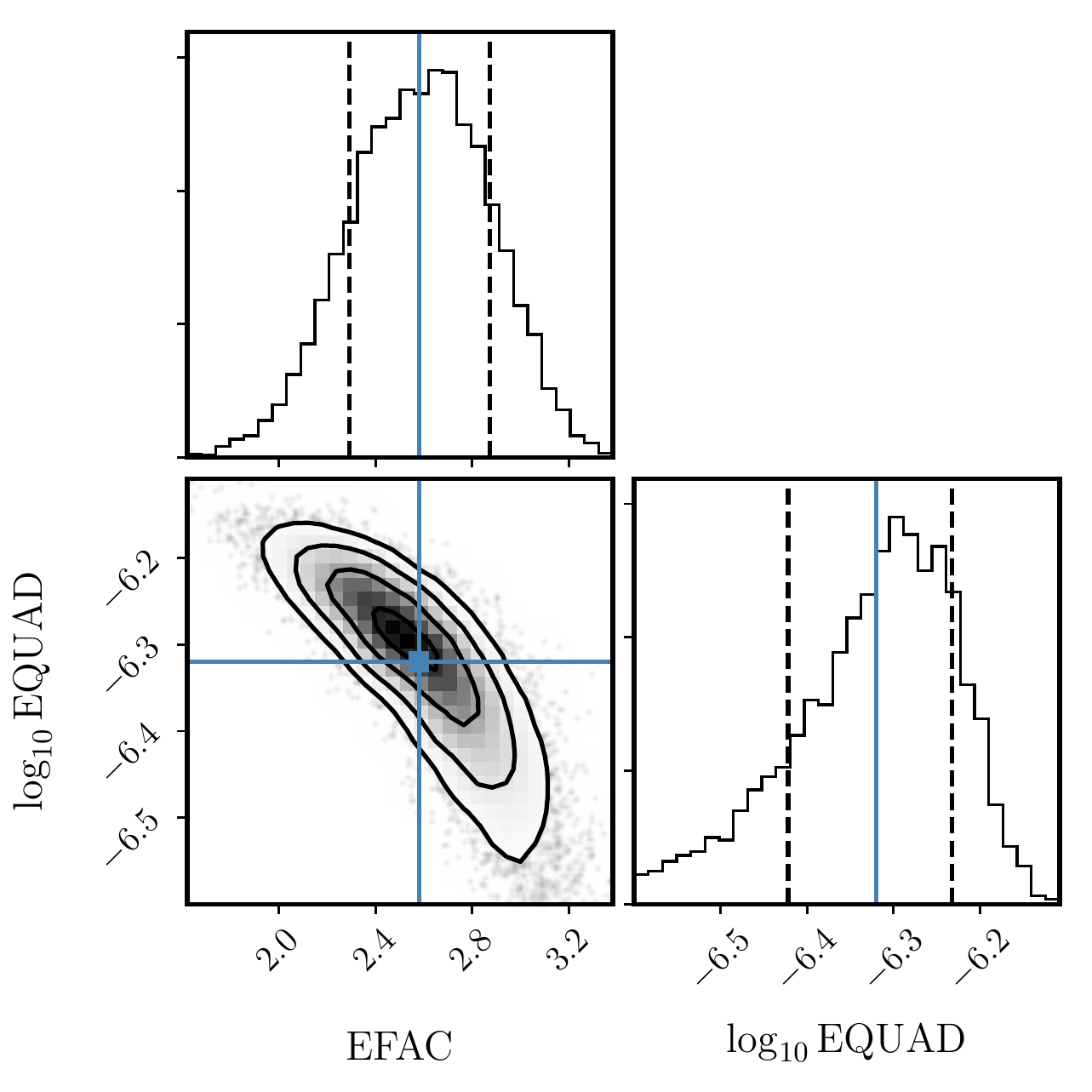}
    \caption{White noise \texttt{ENTERPRISE} timing analysis for J0437$-$4715
    using the A1+A2 data set.}
    \label{fig:enterprise_wn_J04}
\end{figure}

Second, we repeat the previous analysis, now taking both EFAC and EQUAD as free parameters. By doing so, we obtain $\mathrm{EFAC} = 2.48^{+0.29}_{-0.30}$ and $\log_{10}{\mathrm{EQUAD}} = -6.30^{+0.10}_{-0.07}$ ($\mathrm{EQUAD} \approx 0.5~\mu$s) as the best-fit parameters. The quoted error bars correspond to the  1$\sigma$ ($\approx 68\%$) confidence limits that were obtained using the lower-level function \texttt{corner.quantile} from the \texttt{corner.py} Python module \citep{corner} and taking the 16th and 84th percentiles. We present a corner plot for these parameters in Fig.~\ref{fig:enterprise_wn_J04}. In this plot we also show confidence intervals considering that the relevant 1$\sigma$ contour level for a 2D histogram of samples is $1-e
^{-0.5} \sim 0.393$ ($39.3\%$). Values of $\mathrm{EFAC} \sim 1$ would suggest that observing and timing procedures result in near-true TOA uncertainty estimates; thus, the adjusted values of $\mathrm{EFAC} \sim 2.5$ indicate that the TOAs error bars are considerably underestimated.
%

\subsection{Red-noise analysis \label{sec:red_noise}}

The red noise is assumed to be a stationary Gaussian process, which is parameterized with a power-law model in frequency, such that the spectral power density is given by
 \citep[see][]{Hazboun2020ApJ}

\begin{equation}\label{eq:red_noise}
P(f) = \frac{A_\mathrm{rn}^{2}}{12\pi^{2}} \left( \frac{f}{f_\mathrm{ref}} \right)^{\Gamma_\mathrm{rn}}~\mathrm{yr}^{3},
\end{equation}
\noindent where $f$ is a given Fourier frequency in the power spectrum, $f_\mathrm{ref}$ is a reference frequency (in this case, $1 ~ \mathrm{yr}^{-1}$), $A_\mathrm{rn}$ is the amplitude of the red noise at the frequency $f_\mathrm{ref}$, and $\Gamma_\mathrm{rn}$ is the spectral index. We take a prior on the red-noise amplitude that is uniform on $\log_{10} (A_\mathrm{rn}~[\mathrm{yr}^{3/2}]) \in [-14.5,-12]$, and a prior on the red-noise index that is 
uniform on $\Gamma_\mathrm{rn} \in [0,2.6]$. The spectrum is evaluated at 30 linearly spaced frequencies $f \in [1/T_\mathrm{span},30/T_\mathrm{span}]$, where $T_\mathrm{span}$ is the span of the pulsar's data set (in this case, 1.1~yr).

In the following analysis we use a total of 319 observations obtained with A1 and A2 between 2019 April and 2020 June that meet the criteria $t_\mathrm{obs}>40$~minutes, S/N$>40$, and $\sigma_\mathrm{TOA}<1~\mathrm{\mu}$s. Details of this data set are summarized in Table~\ref{table:nobs_sn}.

\begin{figure*}[h]
    \centering
    \includegraphics[width=\linewidth]{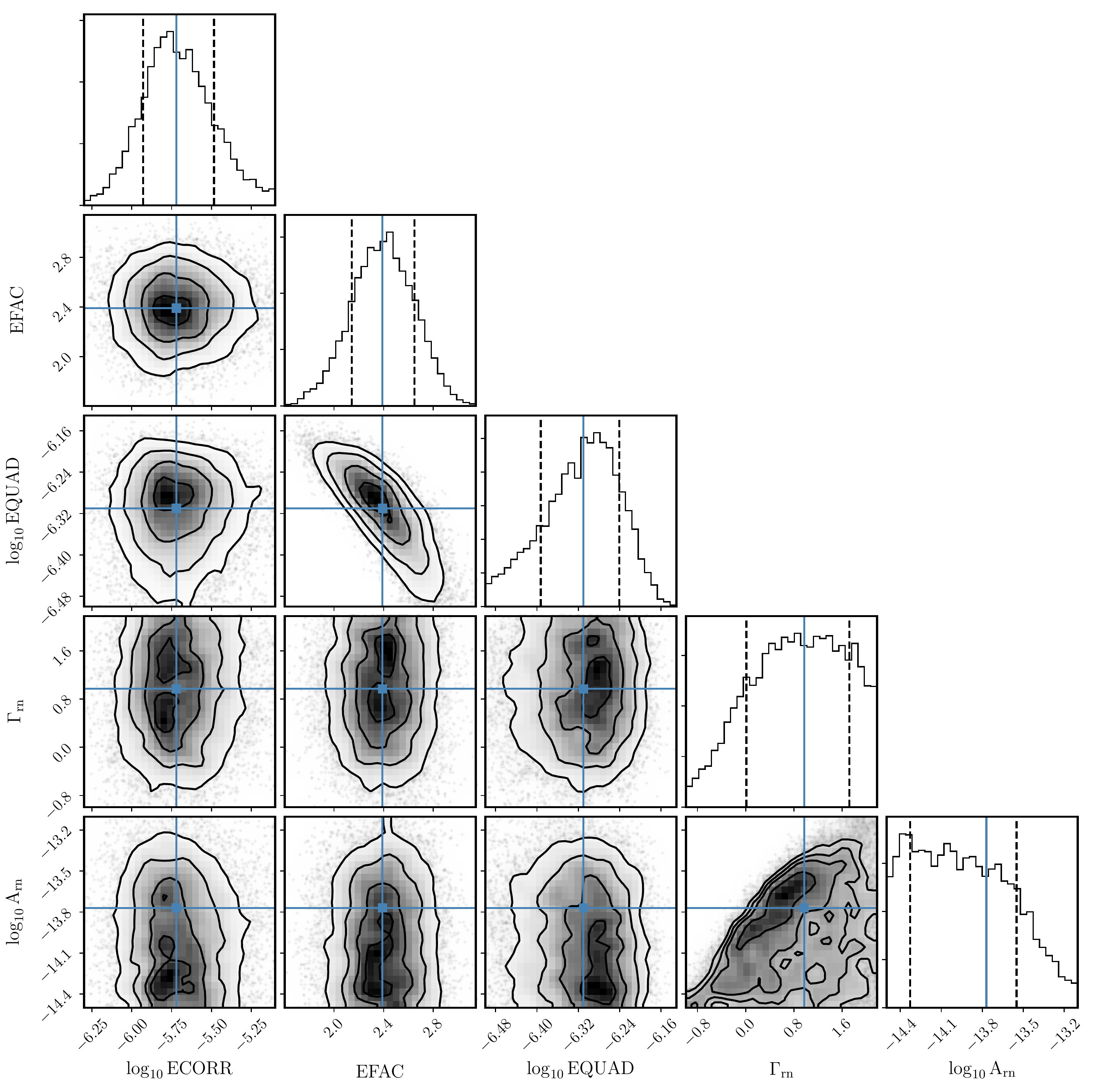}
    \caption{\texttt{ENTERPRISE} timing analysis of red noise for 1.1~yr of observations of J0437$-$4715 (A1+A2).}
    \label{fig:enterprise_J04}
\end{figure*}

We analyze the data sets of each antenna both independently and altogether. As described in Sec.~\ref{sec:white_noise}, ECORR accounts for noise that is correlated between observations that were obtained simultaneously at different frequencies. Since such observations are not available for a single antenna, we exclude the ECORR parameter from the analysis of the individual data sets. However, we do include this parameter when analyzing  the A1+A2 data set in order to profit from the simultaneous observations at different frequencies. A corner plot for these parameters and their errors is shown in Fig.~\ref{fig:enterprise_J04}. 
 
The fitted values to the white- and red-noise parameters for the different data sets are presented in Table~\ref{table:enterprise_J04}.  To complement this we explore the possibility of splitting long-duration observations into two subintegrations of $t_\mathrm{min} = 75$~min in order to sample shorter timing frequencies. The adjusted values for this case, also presented in  Table~\ref{table:enterprise_J04}, are consistent within $1\sigma$ to those obtained without the splitting. We therefore conclude that splitting long observations does not improve the timing analysis, in line with the conclusion from Sect.~\ref{sec:timingsplit}.

We obtain $\mathrm{EQUAD} \approx 0.5~\mu$s in all cases. The value of $\Gamma_\mathrm{rn}$ is less constrained and consistent within 0.5--1.5, while the amplitude is $A_\mathrm{rn}~\approx~2$--$7 \times 10^{-13}$. The obtained $A_\mathrm{rn}$ lies within the expected order of magnitude, whereas $\Gamma_\mathrm{rn}$ falls below the expected value by at least a factor two \citep{Wang:2015bsa}, which we interpret as due to the relatively short baseline of our current data set.
\begin{table}[h]
    {\centering
    \caption{Adjusted values for the noise parameters.}    \label{table:enterprise_J04}
    \begin{tabular}{ l | c c c c }
      & EFAC & $\log_{10} \mathrm{EQUAD}$ & $\Gamma_\mathrm{rn}$ & $\log_{10} A_\mathrm{rn}$ \\
    \hline
    A1 & $2.43^{+0.25}_{-0.23}$ & $-6.3^{+0.06}_{-0.06}$ & $1.22^{+0.13}_{-0.48}$ & $-13.88^{+0.40}_{-0.44}$ \\
    A2 & $2.82^{+0.32}_{-0.30}$ & $-6.34^{+0.08}_{-0.09}$ & $1.02^{+0.36}_{-0.42}$ & $-13.51^{+0.20}_{-0.26}$ \\
    A1+A2 & $2.48^{+0.26}_{-0.24}$ & $-6.32^{+0.09}_{-0.07}$ & $0.97^{+0.37}_{-0.38}$ & $-13.63^{+0.21}_{-0.27}$\\
    A1+A2* & $2.76^{+0.24}_{-0.19}$ & $-6.47^{+0.17}_{-0.14}$ &
    $0.80^{+0.39}_{-0.37}$ &
    $-13.54^{+0.43}_{-0.29}$ \\
    \end{tabular}
    \par}
    \textbf{Note.} Values marked with (*) were obtained by splitting the observations as described in Sec.~\ref{sec:timingsplit}.
\end{table}

\vspace{10mm}

\subsection{GW Analysis \label{sec:gw}}

We now embark on setting the first bounds to the GW amplitude from massive binary black holes using observations from IAR. In doing so, we aim to exploit the high cadence of these observations. 

\subsubsection{Gravitational wave analysis: stochastic background \label{sec:gw_sb}}

The contribution of the GWB coming from an ensemble of supermassive black hole (SMBH) binaries or primordial fluctuations during the big bang is modeled similarly to that of the red noise (Eq.~\ref{eq:red_noise}). Any GWB component is modeled as a single stationary Gaussian 
process with a power-law timing-residual spectral density
\begin{equation}\label{eq:gwb}
    P(f) = \frac{A_\mathrm{gwb}^{2}}{12\pi^{2}} \left( \frac{f}{f_\mathrm{ref}} \right)^{\Gamma_\mathrm{gwb}}~\mathrm{yr}^{3}.
\end{equation}

The analysis is nearly identical to the red-noise analysis described in Sec.~\ref{sec:red_noise}. 
The prior on the GWB amplitude is taken uniform on $\log_{10} (A_\mathrm{gwb}~[\mathrm{yr}^{3/2}]) \in [-14.4,-11]$, whereas the prior on the GWB index is uniform on $\Gamma_\mathrm{gwb} \in [0,3.2]$. Moreover, we fix EFAC and EQUAD to the values adjusted in Sec.~\ref{sec:red_noise} for each data set. 

In this analysis we also consider both the original data sets and the data sets obtained by splitting the observations in subintegrations with $t_\mathrm{obs} \ge 75$ minutes (see Sec.~\ref{sec:timingsplit}). The best-fit values to each GWB parameter and for each set of observations are presented in Table~\ref{table:enterprise_gwb_J04_splitted}.
\begin{table*}[h]
    \centering
    \begin{tabular}{ c c c c c c c  }
    \hline \hline
    Parameter   & \multicolumn{2}{c}{A1}    & \multicolumn{2}{c}{A2}    & \multicolumn{2}{c}{A1+A2}\\
                & no split & split          & no split & split          & no split & split \\
    \hline
    $\Gamma_\mathrm{gwb}$       & $0.50^{+0.25}_{-0.26}$   & $0.38^{+0.20}_{-0.21}$     & $0.12^{+0.04}_{-0.04}$     & $0.10^{+0.03}_{-0.03}$     & $0.38^{+0.28}_{-0.29}$     & $0.28^{+0.17}_{-0.18}$  \\
    $\log_{10} A_\mathrm{gwb}$  & $-13.48^{+0.25}_{-0.23}$      & $-13.37^{+0.20}_{-0.20}$     & $-13.33^{+0.23}_{-0.21}$     & $-13.22^{+0.18}_{-0.17}$      & $-13.48^{+0.24}_{-0.23}$     &  $-13.41^{+0.18}_{-0.18}$  \\
    \hline
    \end{tabular}
    \caption{Best-fit values to the GWB parameters.}
    \label{table:enterprise_gwb_J04_splitted}
\end{table*}
%

Using the split observations, we get a higher cadence at the cost of worsening the S/N (and therefore the TOA precision) of each data point.
Our results show consistent values of $A_\mathrm{gw} \approx (3\pm2) \times 10^{-14}$ and $\Gamma_\mathrm{gw} \approx 0.3\pm0.2$ for all data sets, both with and without the splitting. 
In general, splitting the observations leads to slightly lower values of $\Gamma_\mathrm{gw}$ and slightly higher values of $A_\mathrm{gw}$, though these differences are not significant as they are within $1\sigma$ of the values obtained without the splitting. 

While the amplitude we find is consistent with expected bounds for the stochastic background,  $\Gamma_\mathrm{gw}$ falls short from the expected 13/3 for a stochastic GW background of SMBH binaries \citep{Siemens:2013zla}, possibly due to our current relatively short observational baseline of $\approx 1.1$~yr.

\begin{figure}[htbp]
    \centering
    \includegraphics[width=\linewidth]{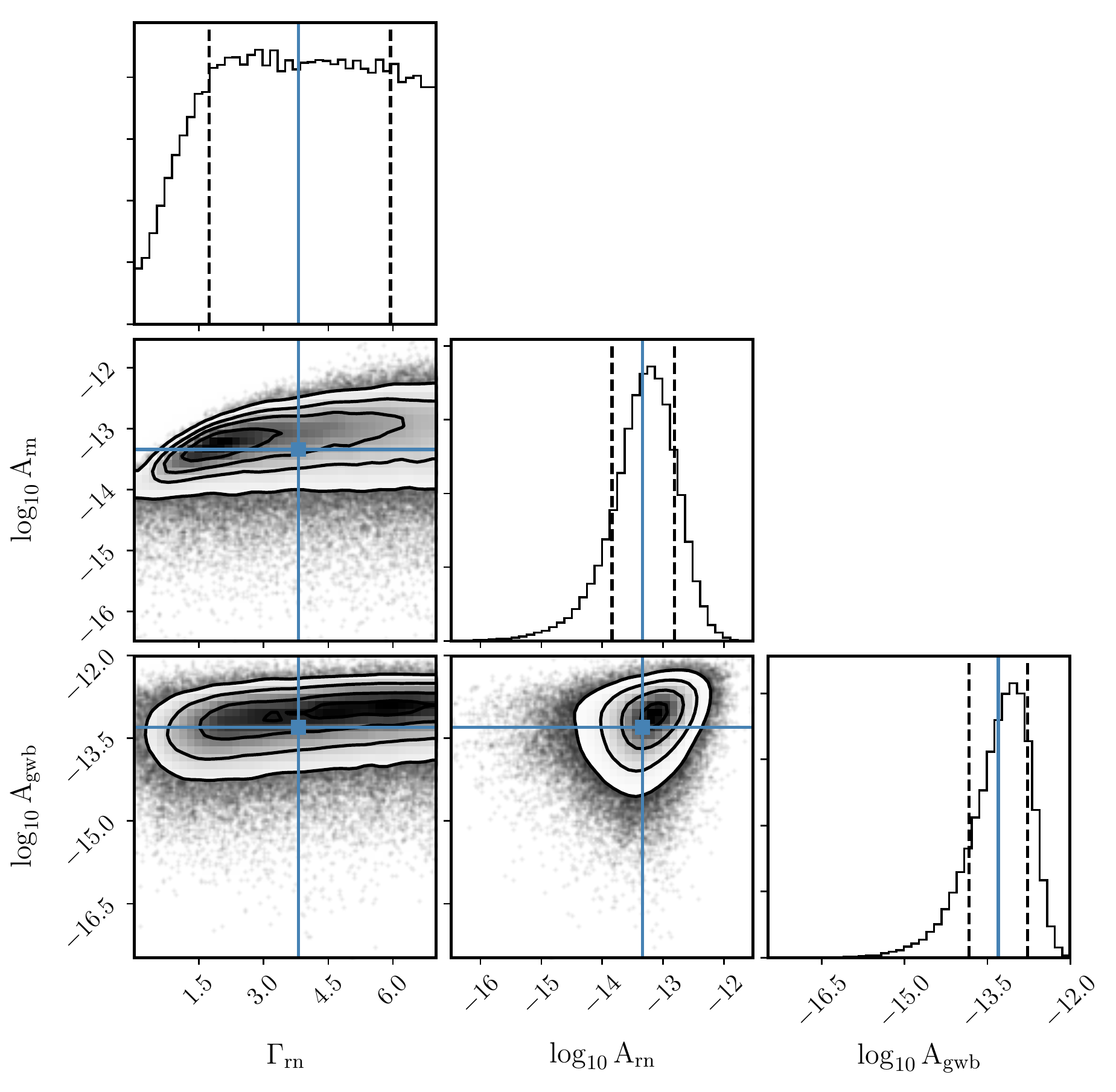}
    \caption{\texttt{ENTERPRISE} gravitational wave analysis for J0437$-$4715 (A1+A2) including a red-noise process with $\Gamma_\mathrm{gwb}=4.33$.}
    \label{fig:enterprise_gwb_final_J04}
\end{figure}

In order to account for a background of SMBH binaries, we repeat this analysis including a red-noise model with a uniform prior on the spectral index $\Gamma_\mathrm{rn} \in [0,7]$ and an extra red-noise process with $\Gamma_\mathrm{gwb}$ set to $4.33$. We also fix all of the white-noise parameters to the values obtained in Sec.~\ref{sec:white_noise}. In Fig.~\ref{fig:enterprise_gwb_final_J04} we show a corner plot of the fit to the joint A1+A2 data sets. We find values of $A_{\mathrm{rn}} \approx (4\pm3) \times 10^{-14}$, consistent with our previous results, and $\Gamma_\mathrm{rn} \approx 3.81 \pm 2.1$. Such uncertainties may be attributed to the short time span.

\subsubsection{GW analysis: continuous source \label{sec:gw_cs}}

A single SMBH binary black-hole system produces “continuous” GWs because the system does not evolve notably over the few years of a pulsar-timing data set. We used the Python package \texttt{Hasasia} \citep{Hazboun2019Hasasia} to calculate the single-pulsar sensitivity curve of our data set of J0437$-$4715 for detecting a deterministic GW source averaged over its initial phase, inclination, and sky location. The dimensionless characteristic strain is calculated for each sampled frequency as
\begin{equation}
    h_{c} (f) = \sqrt{f~S(f)}
\end{equation}
where $S$ is the strain-noise power spectral density for the pulsar. This is related to the power spectrum of the induced timing residuals of Eq.~\ref{eq:gwb} by~\citep[see][]{Jenet:2006sv}
\begin{equation}
    P(f) = \frac{1}{12\pi^{2}} \frac{1}{f^{3}} h_{c}(f)^{2}
\end{equation}

The white- and red-noise parameters that were adjusted in Secs.~\ref{sec:white_noise} and \ref{sec:red_noise} using \texttt{ENTERPRISE} are loaded into the package in order to account for these effects in the calculations. Since our observations have a time baseline of $T_\mathrm{obs}=1.1$ yr and a nearly daily cadence, we calculate the curve across a frequency range between $1/(10~T_{\mathrm{obs}}) \sim 2.8 \times 10^{-9}$~Hz and $1/(1~\mathrm{day}) \sim 1.2 \times 10^{-5}$~Hz.

\begin{figure}[htb]
    \centering
    \includegraphics[width=\linewidth]{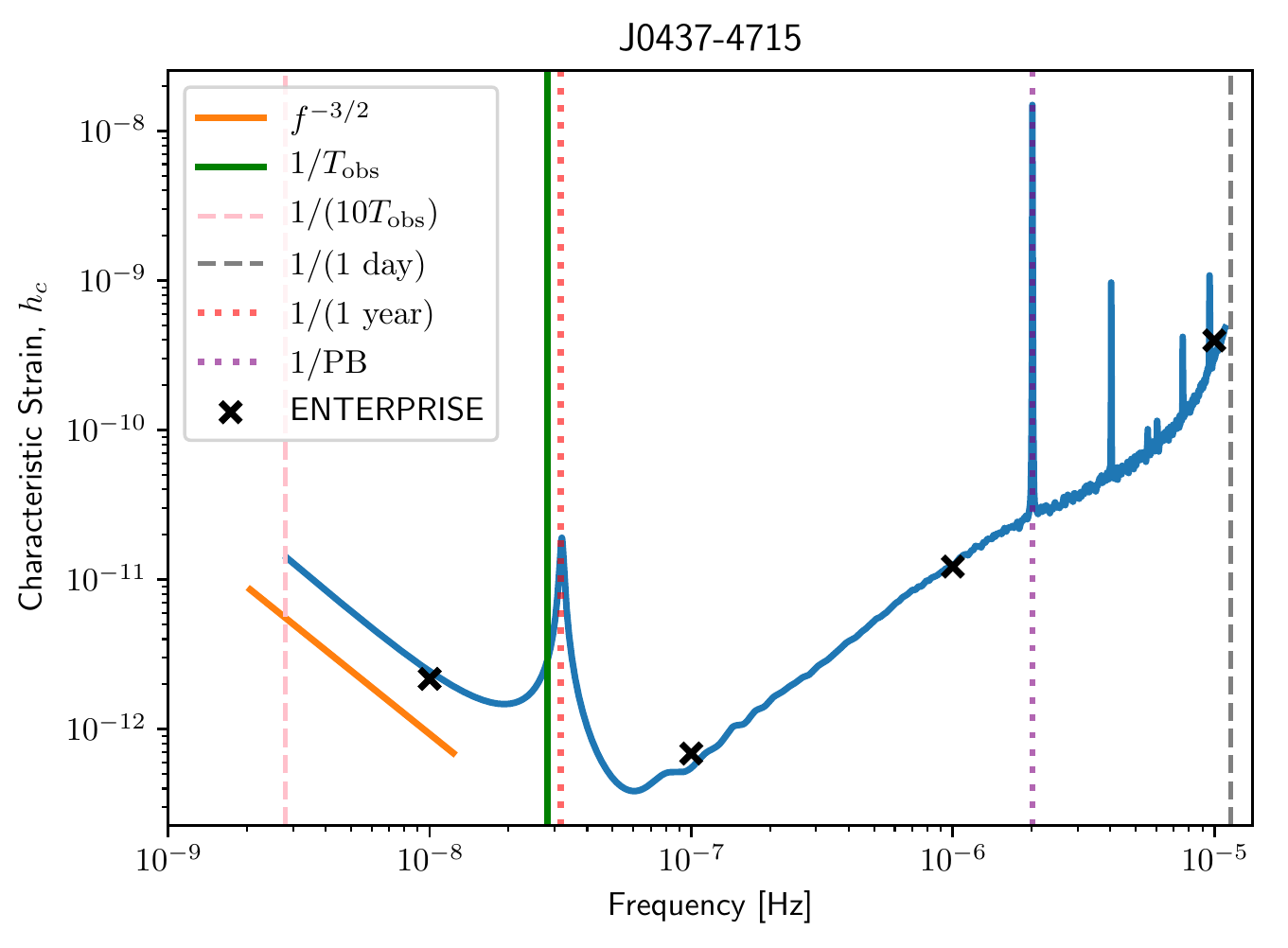}
    \caption{ 
    Sensitivity curve for J0437$-$4715 using 1.1~yr observations at IAR, including pulsar noise characteristics, for a single deterministic gravitational wave source averaged over its initial phase, inclination, and sky location (A1+A2; blue curve). The vertical green line corresponds to a frequency of $1/T_{\mathrm{obs}}$, the dotted red line to $1/T_{\mathrm{yr}}$ and the dotted purple line to $1/{\mathrm{PB}}$ (orbital period). The black crosses correspond to the mean values of the $\log_{10} h_\mathrm{gw}$ distributions obtained using \texttt{ENTERPRISE}.}
    \label{fig:sensitivity_J04}
\end{figure}

The resulting sensitivity curve is shown in Fig.~{\ref{fig:sensitivity_J04}}. It is readily seen that there is a loss of sensitivity at a frequency of $(1~\mathrm{yr})^{-1}$, caused by fitting the pulsar's position, and at a frequency of $(\mathrm{PB})^{-1} \sim 2~\mathrm{\mu}$Hz (with $\mathrm{PB}$ the orbital period), caused by fitting the orbital parameters of the binary system. The additional spikes seen at frequencies higher than $(\mathrm{PB})^{-1}$ correspond to harmonics of the binary orbital frequency.

In addition, the sensitivity at lower frequencies is reduced by (i) the fit of a quadratic polynomial to the TOAs required to model the pulsar spin-down and (ii) the fitting of `jumps' to connect the timing residuals obtained with different backends \citep{Yardley2010}. The frequency dependence ($\sim f^{-3/2}$) at low frequencies is evidence of a fit to a quadratic spin-down model for the pulsar spin frequency. As a result, the minimum of the sensitivity curve should be attained at a frequency of $1/T_\mathrm{obs}$. However, given that the $T_\mathrm{obs}$ of our data set is close to 1 yr, this feature coincides with the loss of sensitivity at $(1~\mathrm{yr})^{-1}$. We expect to obtain a well-defined minimum at $\approx 1/T_\mathrm{obs}$ in a future by accumulating more observations and achieving a significantly longer time baseline.

For completeness, we tested the significance of the red-noise contribution by calculating a sensitivity curve without this component. The curve was essentially insensitive to those changes in the priors. This is expected, since the injection of red noise should lead to a flat sensitivity curve around the minimum \citep{Hazboun2019}, though in our case it is coincident with the spike at $(1~\mathrm{yr})^{-1}$.


For comparison, we used \texttt{ENTERPRISE} to perform a fixed-frequency Markov chain Monte Carlo procedure at four different frequencies. We obtained a posterior distribution for $\log_{10} h_\mathrm{gw}$ at each of these frequencies with a mean value in great agreement with the curve obtained with \texttt{Hasasia}, as shown in Fig.~\ref{fig:sensitivity_J04}. 


%
These first results on GW sensitivity are encouraging, though we still need to achieve a sensitivity of at least a factor 10 higher in order to observe even the most favorable SMBH binary merger events. For instance, the six billion solar mass source of 3C~186 at $z\approx1$ produced a GW of $h\sim 10^{-14}$ at the time of arrival to our Galaxy, roughly a million years ago~\citep{Lousto:2017uav}.

%
\section{Conclusions}\label{sec:conclusions}
%

We presented the first detailed analysis of the observational campaign toward the bright MSP J0437$-$4715 using the two antennas at IAR's observatory. This data set comprises over a year of high-cadence (up to daily) observations with both antennas, A1 and A2.

We quantified the timing precision and noise parameters using the current setup for A1 and A2. We also explored the effect of different reduction parameters of the raw data. We conclude that as follows:
\begin{itemize}
    \item The number of phase bins used in the reduction does not have an impact on the timing precision as long as $n_\mathrm{bins}\geq 256$.
    \item The S/N of the individual observations plays a crucial role in determining the timing precision. In particular, to achieve a timing precision $< 1~\mu$s, observations with $\mathrm{S/N} > 140$ are required, a condition that is currently fulfilled by $\sim1/3$ of the observations taken with A2 and $\sim1/2$ of the observations taken with A1.
    \item Splitting long observations into shorter intervals does not improve the timing precision, most likely due to current limitations in the S/N for short observations.
    \item A1 slightly outperforms A2, probably 
    due to its larger bandwidth configuration. 
    \item The systematic errors of the observations are $\sigma_\mathrm{sys} \approx 0.5~\mu$s, although this value is likely to be S/N-limited. The rms of the data set is $\approx 0.5$--0.6$~\mu$s
    \item The white-noise analysis performed with \texttt{ENTERPRISE} indicates that the error bars are typically underestimated by a factor $\sim$3 when accounting for EQUAD and EFAC.
    \item We placed upper limits to the GWB in the tens of nHz to sub-$\mu$Hz frequency range. Although the current sensitivity is not sufficient for placing physically interesting constraints, the ongoing campaign--together with incoming hardware upgrades--is likely to significantly improve in the next five to ten years \citep[see also][]{Lam2020}. In particular, observations lasting over 3~hr are promising for exploring GW signals with frequencies above 0.1~$\mu$Hz by splitting them into hour-scale subintegrations.
\end{itemize} 

Ongoing and future hardware upgrade of IAR's antennas, such as installing larger-bandwidth boards, promise to expand IAR's observational capabilities and improve its achievable timing precision. Such upgrades would allow us to reduce the systematical errors of the antennas and to include (sub)daily high-precision timing of other MSPs of interest, such as PSR J2241$-$5236.

\acknowledgments

The authors thank various members of the IAR's technical staff and board for their work and support, as well as numerous members of the LIGO-Virgo and NANOGrav collaborations for very valuable discussions, in particular P. Baker, J. Hazboun, M. Lam, and S. Taylor.
Part of this work was supported by the National Science Foundation (NSF) from Grants No. PHY-1912632, No.\ PHY-1607520, and No.\ PHY-1726215. V.S.F. acknowledges financial support from a summer fellowship of the Asociaci\'on Argentina de Astronom\'ia (AAA). F.G. and J.A.C. acknowledge support by PIP 0102 (CONICET). 
This work received financial support from PICT-2017-2865 (ANPCyT). J.A.C. was also supported by grant PID2019-105510GB-C32/AEI/10.13039/501100011033 from the Agencia Estatal de Investigaci\'on of the Spanish Ministerio de Ciencia, Innovaci\'on y Universidades, and by Consejer\'{\i}a de Econom\'{\i}a, Innovaci\'on, Ciencia y Empleo of Junta de Andaluc\'{\i}a as research group FQM- 322, as well as FEDER funds.

%

\vspace{5mm}
\facility{IAR}



\software{\texttt{ENTERPRISE} \citep{Ellis2019}, 
        \texttt{PRESTO} \citep{Ransom2003, Ransom2011}, 
        \texttt{psrchive} \citep{Hotan:2004tz}, 
        \texttt{PyPulse} \citep{PyPulse}, 
        \texttt{Hasasia} \citep{Hazboun2019Hasasia}
}



\appendix

\section{Details of the analysis}
\label{sec:appendices}

\subsection{Reduction of observations\label{sec:reduction}}

To de-disperse and fold the observations we use the software \texttt{PRESTO} \citep{Ransom2003, Ransom2011}. It has a variety of tools for the reduction of observations.
The processed data are stored in a \texttt{.pfd} file that contains the pulse profile for different time and frequency bins. In addition to this profile, \texttt{PRESTO} outputs a \texttt{.polycos} file that contains the coefficients of a polynomial modeling the variation of the pulsar period. These coefficients allow us to determine the period of pulsation in a topocentric reference system and are necessary to compute the timing residuals.

If the observation is in the file \texttt{obs.fil} and the mask in the file \texttt{m.mask}, then the command-line used has the following syntax:
\begin{multline}
    \texttt{prepfold -nsub 64 -n 1024 -timing} \\
    \hfill \texttt{J0437$-$4715.par -mask m.mask obs.fil}
\end{multline}
\noindent 
where the option \texttt{-timing} indicates \texttt{prepfold} to generate a file \texttt{.polycos} based on the pulsar parameters 
This process is currently automatized through local Python scripts. 

\subsection{Templates} \label{sec:templates}

Considering that A1 and A2 have different configurations (number of polarizations and BW; see Table~\ref{table:par_obs}), it is possible that slight differences arise in the integrated profile seen by each antenna. We therefore study whether the template used has a significant impact on the timing residuals.

\begin{figure}[h]
    \centering
    \includegraphics[width=\linewidth]{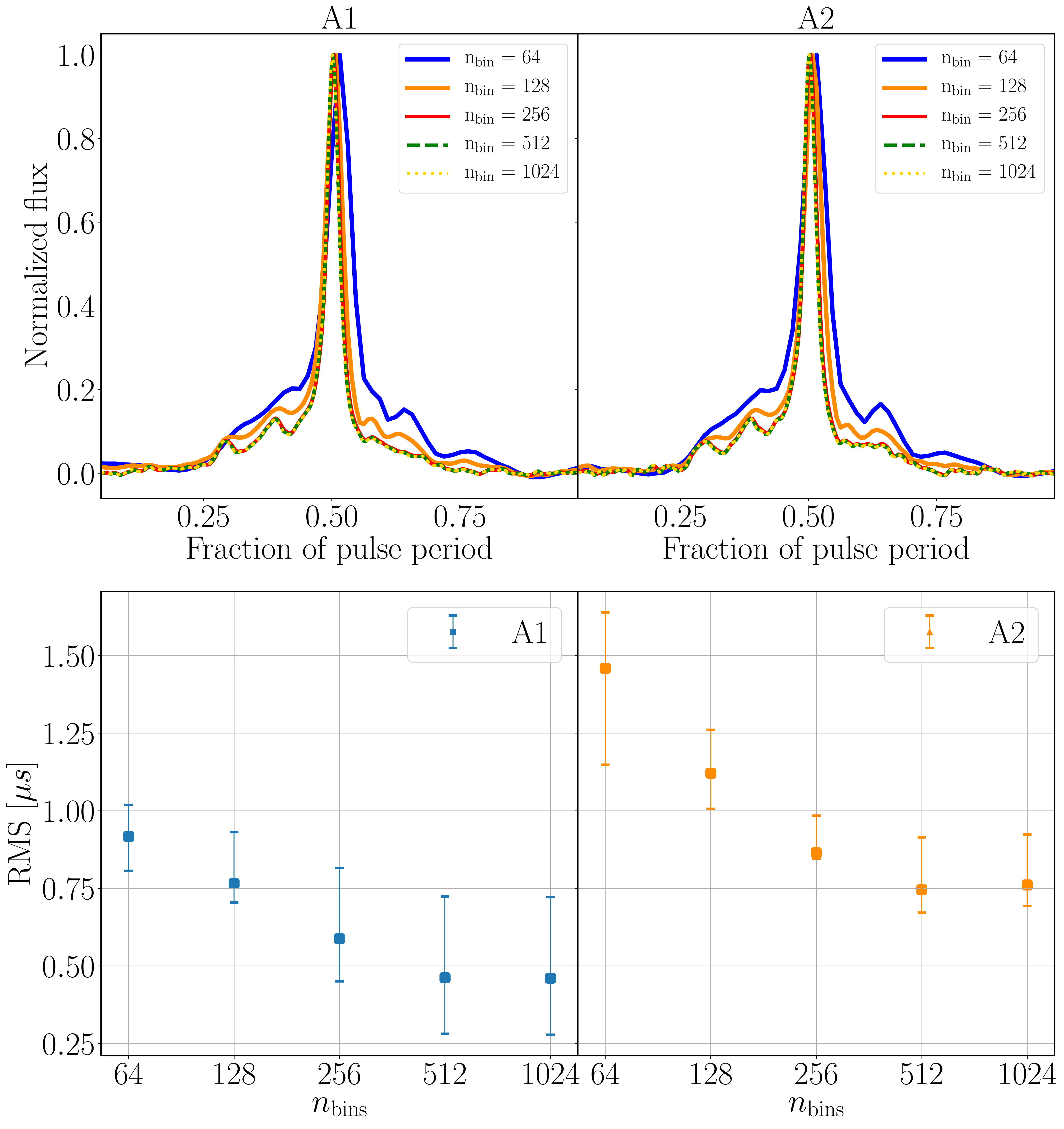}\\
    \caption{Top: \textit{emplates} for each antenna for different values of $n_\mathrm{bins}$. Bottom: rms found for each subset per $n_\mathrm{bins}$, and its corresponding 1$\sigma$ error bars. }
    \label{fig:nbins_templates_rms}
\end{figure}

To create each template we choose observations with $n_\mathrm{bins}=1024$ phase bins and $n_\mathrm{chan}=64$ frequency channels. We select for each antenna data the highest-S/N observation and extract the noise by using the task \texttt{psrsmooth} in the package \texttt{psrchive}. This choice of templates seems adequate since J0437$-$4715 is a very bright pulsar and selected individual observations produce a high enough S/N to create a template.
We highlight that the large span of our observations (over 3 hr in many cases) mitigates the impact of the intrinsic jitter of the pulsar \citep{Liu:2011te}.
The selected templates for each antenna correspond to the profiles with $n_\mathrm{bins}=1024$ phase bins in Fig.~\ref{fig:nbins_templates_rms}. The relative error between them is below $5\%$ near the peak, with larger relative differences toward the wings, but those do not have major influence in the determination of the TOAs. 

In the preliminary timing analysis of J0437$-$4715 presented in \cite{Gancio2020}, we used the same template on both antennas to determine TOAs. We show in our separated analysis of A1 and A2 data that this assumption was valid to the current level accuracy, producing an $\mathrm{rms}~=~0.8~\mathrm{\mu s}$ residual for A2 observations with the use of either template. Notwithstanding this {\it a posteriori} verification, we consistently use different templates for A1 and A2 throughout this work.

%

\subsection{Timing versus number of phase bins \label{sec:timingnbins} }

In order to study the effect of the number of phase bins ($n_\mathrm{bins}$) used in the folding of observations on the timing residuals, we have taken data folded originally with $n_\mathrm{bins}=1024$, and processed with the routine \texttt{bscrunch} of the \texttt{psrchive} package for Python, to generated copies of the observations and their corresponding templates for each antenna, but with values of $n_\mathrm{bins}$= 512, 256, 128, 64, and 32. Through this process of
scrunching we obtained six sets of observations for each antenna only differing by their $n_\mathrm{bins}$.
 Fig.~\ref{fig:nbins_templates_rms} shows the effect of the $n_\mathrm{bins}$ on the 
 templates for each antenna. While for $n_\mathrm{bins}=32$ we lose temporal resolution, the differences beyond $n_\mathrm{bins} \ge 256$ are almost negligible to our precision.
Next we compute the timing residuals for each $n_\mathrm{bins}$ subset.
Interestingly, only for $n_\mathrm{bins} \le 64$ are the timing errors are too large; 
for $n_\mathrm{bins} \geq 128$ the derived TOAs are very consistent, the size of the error bar being the main difference (with smaller error bars obtained for larger $n_\mathrm{bins}$). 
%
%
The rms of the residuals for each subset after adjusting $\sigma_\mathrm{sys}$ as a function of $n_\mathrm{bins}$ is shown in Fig.~\ref{fig:nbins_templates_rms}. The rms decreases with increasing $n_\mathrm{bins}$ significantly for $64$ to $256$ bins, showing that we cannot attain good timing for $n_\mathrm{bins} \le 64$ and need at least 256 bins to obtain a precision higher than $1 \mathrm{\mu s}$ for a pulsar like J0437$-$4715. This corresponds to a time interval much smaller than the full width at half-maximum of the pulse, that is,  0.3~$\mu$s at 1400 MHz. 
%

\subsection{Timing versus S/N\label{sec:timing_vs_sn}} 
%
\begin{figure*}[htb!]
  \includegraphics[width=\columnwidth]{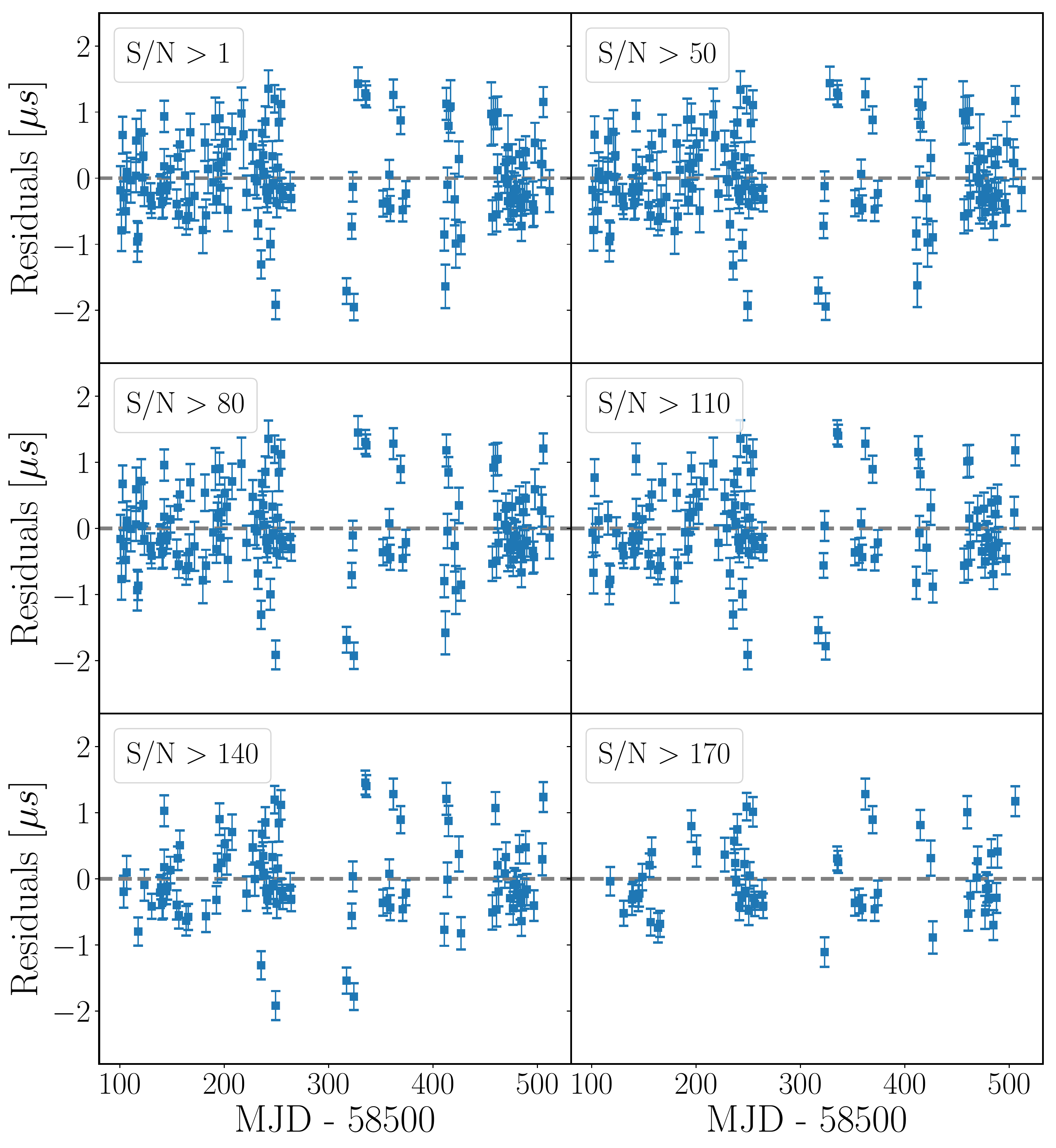} 
  \includegraphics[width=\columnwidth]{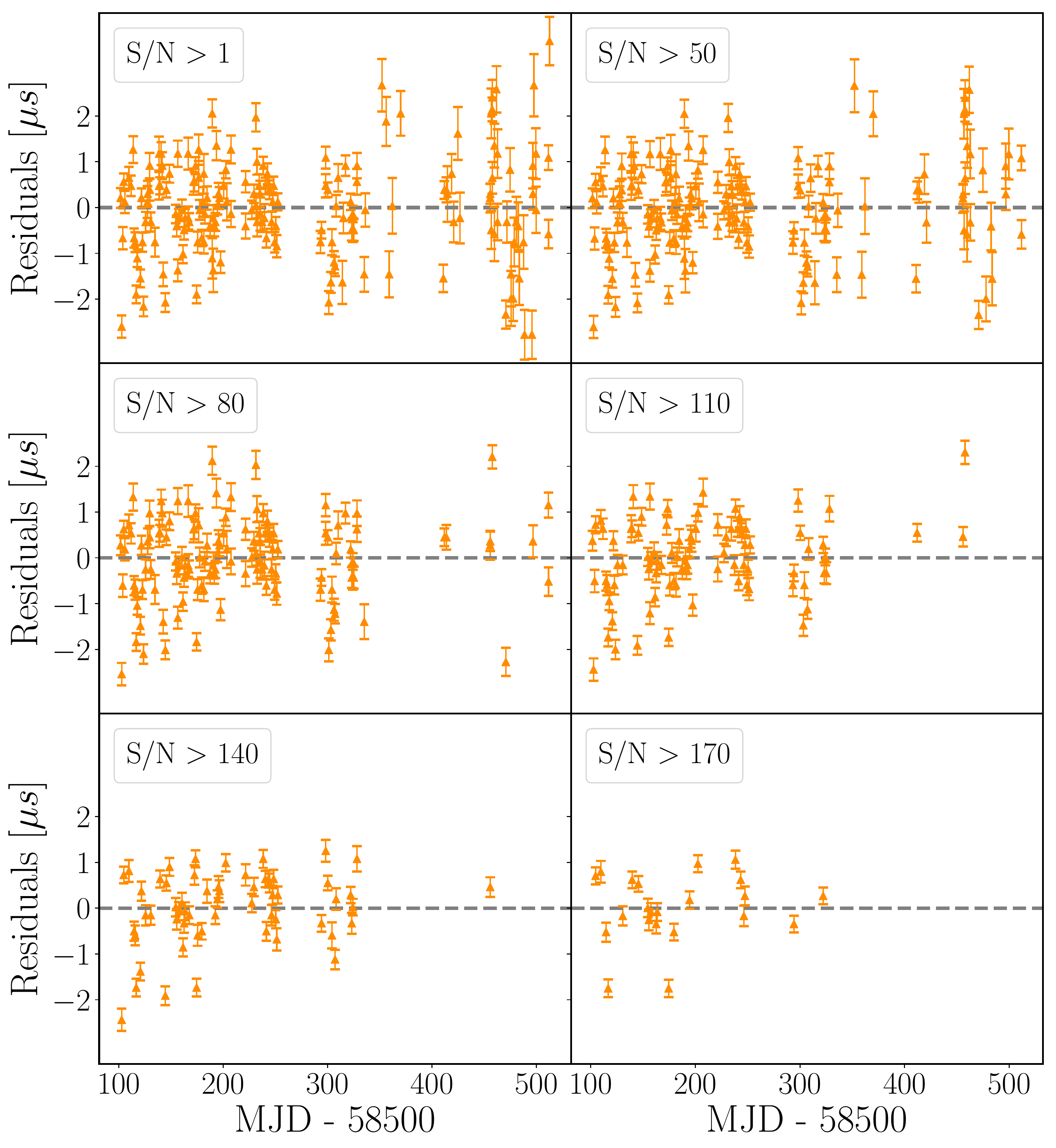}%
  \caption{Residuals of each subset of observations with A1 (left) and A2 (right) grouped in different data sets according to their minimum S/N (see legends). }
  \label{fig:residues_sn_A1_A2}
\end{figure*}

Here we present additional figures that support the hypothesis that our timing studies are limited due to the S/N of the observations. This effect has a larger impact for A2, as can be seen in Fig.~\ref{fig:residues_sn_A1_A2}.



\bibliography{biblio}{}
\bibliographystyle{aasjournal}



\end{document}